\documentclass[12pt,a4paper]{article}

\usepackage[utf8]{inputenc}
\usepackage[T1]{fontenc}
\usepackage{lmodern}
\usepackage[margin=2.5cm]{geometry}
\usepackage{microtype}

\usepackage{amsmath,amssymb,amsthm,mathtools}
\usepackage{bm}

\usepackage{float}
\usepackage{booktabs,array,multirow,threeparttable}
\usepackage{graphicx}
\usepackage{xcolor}

\usepackage[most]{tcolorbox}

\usepackage[authoryear,round]{natbib}

\theoremstyle{plain}
\newtheorem{theorem}{Theorem}[section]
\newtheorem{proposition}[theorem]{Proposition}
\newtheorem{lemma}[theorem]{Lemma}

\theoremstyle{definition}

\theoremstyle{remark}
\newtheorem{remark}[theorem]{Remark}

\definecolor{theoremblue}{RGB}{0,51,102}
\definecolor{theoremlightblue}{RGB}{230,240,255}
\definecolor{defgreen}{RGB}{0,80,40}
\definecolor{deflightgreen}{RGB}{230,248,230}
\definecolor{remarkgray}{RGB}{60,60,60}
\definecolor{remarklightgray}{RGB}{240,240,240}

\newtcolorbox{keyresult}[1][]{
  colback=theoremlightblue,
  colframe=theoremblue,
  fonttitle=\bfseries,
  title={#1},
  boxrule=0.8pt,
  arc=2pt,
  left=6pt,
  right=6pt,
  top=4pt,
  bottom=4pt
}

\newtcolorbox{keydef}[1][]{
  colback=deflightgreen,
  colframe=defgreen,
  fonttitle=\bfseries,
  title={#1},
  boxrule=0.8pt,
  arc=2pt,
  left=6pt,
  right=6pt,
  top=4pt,
  bottom=4pt
}

\usepackage[mathlines]{lineno}

\newif\iflinenumbers
\linenumbersfalse

\AtBeginDocument{%
  \iflinenumbers
    \linenumbers
  \fi
}

\usepackage[
  colorlinks=true,
  linkcolor=blue!60!black,
  citecolor=green!50!black,
  urlcolor=blue!70!black
]{hyperref}

\usepackage{cleveref} 
\title{Stationary Analysis of Adaptive Aspiration Dynamics in Graph-Structured Games}
\author{Dhaker Kroumi\\
Department of Mathematics\\
King Fahd University of Petroleum and Minerals,\\
Dhahran, Saudi Arabia\\
dhaker.kroumi@kfupm.edu.sa}
\date{}

\begin{document}

\maketitle

\begin{abstract}
Aspiration dynamics can favor cooperation on finite graphs through a purely
local satisfaction mechanism. In the model studied here, individuals compare
their payoffs with adaptive aspiration levels and revise their strategies
through a self-switching Fermi rule. Although the aspirations are continuous
and endogenous, the resulting finite-population process has a
unique stationary distribution for every finite selection intensity and every
finite connected graph under degree-normalized average payoffs when restricted
to the compact forward-invariant state space in which aspirations lie in the
payoff interval. In the weak-selection regime,
the first-order stationary abundance shift is positive when
\(R+S>T+P\), negative when \(R+S<T+P\), and zero when
\(R+S=T+P\). In the equality case, higher-order terms may determine its
direction. The condition \(R+S>T+P\) means that, under neutrality, the expected payoff of a cooperator exceeds
that of a defector. The stationary variance reveals a different effect whose
first-order correction is governed by the payoff-range contrast
\((R-S)-(T-P)\). This contrast compares how strongly cooperator and defector
payoffs respond to changes in local neighborhood composition and therefore
measures the relative sensitivity of their payoffs to the
local environment. Under degree-normalized average payoffs,
these first-order results hold on every finite connected graph and are
independent of degree sequence, topology, and aspiration learning rate.
Simulations on regular and heterogeneous graphs compare
averaged and accumulated payoffs and illustrate that graph heterogeneity and
the aspiration learning rate can affect stationary behavior away from
weak selection. The first-order coefficients are derived from the finite
stochastic process and do not use pair approximation, mean-field closure, or
a large-population limit.

\end{abstract}

\section{Introduction}

Cooperation among unrelated individuals is a central problem in evolutionary biology, economics, and the social sciences. Costly social behavior poses a challenge for simple individual-level advantage, and several mechanisms have been proposed to explain how cooperation can persist, including kin assortment, reciprocity, population structure, and group interactions \citep{Darwin1859,Hamilton1964,Trivers1971,AxelrodHamilton1981,Axelrod1984,Nowak2006}. Evolutionary game theory provides a general mathematical framework for studying such frequency-dependent social interactions \citep{MaynardSmith1982,HofbauerSigmund1998}.

One classical representation of population structure
partitions the population into relatively discrete groups or demes.
Social interactions occur primarily within demes, whereas migration,
group reproduction, extinction, or recolonization connect the demes at
the population level. Such models provide the basis for group- and
multilevel-selection approaches to cooperation
\citep{MaynardSmith1964GroupSelection,Wilson1975,
TraulsenNowak2006Multilevel,LehmannKellerWestRoze2007}. Within this
framework, differential and sensitivity-driven migration can favor
cooperation when cooperators disperse more readily from defector-rich
demes \citep{KroumiLessard2025,KroumiPRE2025}. Cooperation can also be
promoted when defector-rich demes face a greater probability of
extinction and subsequent recolonization
\citep{KroumiJTB2025}.

A different representation of population structure
arises when individuals interact through local and potentially
overlapping neighborhoods. Early work on spatial games
demonstrated that local interactions allow cooperators to form
protective clusters, helping them resist exploitation by defectors
\citep{NowakMay1992}. This concept was later extended to graphs and
complex networks under the framework of network reciprocity
\citep{SantosPacheco2005}. In particular, weak-selection calculations
have yielded exact mathematical conditions for cooperation under
various update rules
\citep{OhtsukiNowak2006,OhtsukiHauertLiebermanNowak2006}. Subsequent
analyses using finite graphs have highlighted that whether cooperation
succeeds depends strongly on the update rule, the method of payoff
accounting, and the underlying network topology
\citep{TaylorDayWild2007,SzaboFath2007,RocaCuestaSanchez2009,
PercSzolnoki2010}.

In addition to network topology, the scale of interaction plays a major role in these dynamics. While many models focus on pairwise interactions, others look at group-level social dilemmas through public-goods or collective-action problems \citep{Hardin1968,Dawes1980}. Research on nonlinear public goods, volunteering, and snowdrift games confirms that the effect of population structure varies widely depending on the specific game and payoff rules \citep{HauertDeMonteHofbauerSigmund2002,HauertDoebeli2004,SantosSantosPacheco2008,ArchettiScheuring2012}. Consequently, broad reviews emphasize the need to study network structure, stochasticity, and behavioral update rules together \citep{PercJordanRandWangBoccalettiSzolnoki2017}.

Within this framework, classical finite-population models often rely on imitation-driven dynamics to analyze stochastic properties like invasion, fixation, and long-term stability \citep{TaylorFudenbergSasakiNowak2004,TraulsenNowakPacheco2006,FudenbergImhof2006,KandoriMailathRob1993,Young1993}. Under this rule, individuals update their behavior by simply copying the strategies of neighbors who receive higher payoffs. However, imitation is not the only plausible way individuals change their behavior. Insights from reinforcement learning, bounded rationality, and satisficing suggest that individuals often adjust their strategies by comparing their own experienced outcomes against internal standards, rather than by looking at others \citep{BushMosteller1955,Simon1955,Selten1998,ErevRoth1998}. Aspiration-based learning formalizes this principle of self-evaluation.

Under this framework, each individual evaluates its success using its own aspiration level, which serves as an internal performance standard. If the individual is satisfied with its payoff relative to this standard, it usually maintains its strategy; if it is dissatisfied, it becomes more likely to try the alternative behavior. Therefore, strategy revision is a purely private process driven by personal disappointment. This focus on individual experience removes payoff comparison with peers, while retaining social interaction through the payoff structure.

Aspiration dynamics under weak selection were studied in finite well-mixed populations with a common fixed aspiration level by \citet{DuWuAltrockWang2014}, and were later extended to regular graphs by \citet{DuWuWang2015}. Subsequent analytical work considered individualized but constant aspirations on regular networks under weak selection \citep{WuZhou2018}. A related extension studied heterogeneous aspiration-based update functions and established an additivity property under weak selection, with additional results beyond weak selection for public goods games \citep{ZhouWuVasconcelosWang2018}. More recently, \citet{ZhouWuDuWang2021} showed that aspiration dynamics generate robust weak-selection predictions on heterogeneous weighted networks, allowing distributions of aspiration levels and individualized self-evaluation rules. In these works, aspiration levels are treated as fixed parameters, or individual heterogeneity is encoded in fixed aspiration values or update functions, rather than as payoff-dependent state variables that evolve with the population process.

A second branch of literature studies richer behavioral and structural mechanisms in which aspirations, strategies, networks, or decision processes adapt. Examples include coevolutionary aspiration and strategy dynamics in spatial repeated public goods games \citep{WuFuWang2018}, public goods games with payoff aspirations \citep{LiuJiaRong2019}, network rewiring driven by aspirational learning \citep{PalSengupta2022}, aspiration-driven coevolution with individual behavioral diversity \citep{YuanWangWangYangXuHuang2023}, and decision uncertainty in aspiration-driven games on complex networks \citep{AguilarJanitaKhalilLeyvaSendinaNadal2025}. These models capture important behavioral and structural effects, but their added mechanisms create high-dimensional dependencies. As a result, they are typically analyzed through simulation or approximate statistical-physics arguments, rather than through exact finite-population stationary calculations.

Many analytical treatments in this literature use large-population limits, pair approximations, or simulation-based evidence. Here we instead derive the first-order stationary coefficients directly for the finite-graph process. In our model, each individual dynamically updates its aspiration by moving a fixed fraction toward its current payoff, and revises its strategy through an aspiration-based Fermi self-switching rule.  We first prove that when restricted to a compact invariant space, this exact process admits a unique invariant probability measure, ensuring that the stationary state is well-defined. We then derive exact one-step drift identities for the cooperator count and the strategy-specific aspiration sums. These identities allow us to calculate first-order weak-selection expansions not only for the mean stationary abundance, but also for its variance. By analyzing the variance alongside the mean, we can quantify the statistical dispersion of the process and measure how much the strategy frequencies fluctuate around their expected values. Ultimately, our analysis shows that the mean abundance indicates which strategy is more satisfying on average, while the variance captures how much a strategy is influenced by fluctuations in local neighborhood composition.

The paper is organized as follows.  Section~\ref{sec:model} describes the model and establishes the stationary-distribution framework. Section~\ref{sec:exact-drift} derives the exact drift identities alongside the weak-selection formulas for stationary abundance and variance. These formulas are then applied to standard pairwise games in Section~\ref{sec:applications}. To complement the analytical insights, Section~\ref{sec:simulations} compares the theoretical results with Monte Carlo simulations. Finally, Section~\ref{sec:discussion} discusses the biological interpretation, novelty, and limitations of our findings, while Appendix~\ref{appendixA} contains the mathematical proof for the uniqueness of the stationary distribution.
\section{\label{sec:model}Model and Stochastic Process}
Consider a simple connected graph \(G=(V,E)\). The vertex set \(V\) satisfies
\(|V|=N\), and each vertex is occupied by one individual.
\(E\) is the set of edges specifying the interaction network. For two
vertices \(i\) and \(j\), \(j\sim i\) means that vertices \(i\) and
\(j\) are adjacent. For each \(i\in V\), let
\[
d_i=\#\{j\in V:j\sim i\}
\]
denote the degree of vertex \(i\). Throughout the
paper, we assume \(2\le N<\infty\). Since \(G\) is simple and
connected, we have
\[1\le d_i\le N-1,\]
for every $i\in V.$
At time \(t\), individual \(i\) has a strategy
\(\mathcal{S}_i(t)\in\{C,D\}\) and an aspiration level
\(A_i(t)\in\mathbb R\). The strategy \(C\) denotes cooperation and
\(D\) denotes defection. The aspiration level is the payoff standard
against which the individual evaluates its current performance.

The frequency of cooperation is given by
\[
p_C(t)=\frac1N\sum_{i\in V}\mathbf 1_{\{\mathcal{S}_i(t)=C\}},
\]
while the frequency of defection is then
\(p_D(t)=1-p_C(t)\). For each vertex \(i\), let
\[
n_i(t)=\sum_{j\sim i}\mathbf 1_{\{\mathcal{S}_j(t)=C\}}
\]
be the number of cooperative neighbors of \(i\).
Thus \(d_i-n_i(t)\) is the number of its defecting
neighbors.

Individuals interact pairwise with their neighbors according to the
payoff matrix
\[
\begin{array}{c|cc}
 & C & D\\
\hline
C & R & S\\
D & T & P
\end{array}.
\]
A cooperator receives payoff \(R\) against a cooperative neighbor
and payoff \(S\) against a defecting neighbor. A defector receives
payoff \(T\) against a cooperative neighbor and payoff \(P\) against
a defecting neighbor.

The payoff of individual \(i\) is the average payoff over its
\(d_i\) pairwise interactions:
\[
\pi_i(t)=
\begin{cases}
\dfrac{ n_i(t)R+\bigl(d_i-n_i(t)\bigr)S}{d_i}=x_i(t)R+\bigl(1-x_i(t)\bigr)S,& \text{if }\mathcal{S}_i(t)=C,\\[3mm]
\dfrac{ n_i(t)T+\bigl(d_i-n_i(t)\bigr)P}{d_i}=x_i(t)T+\bigl(1-x_i(t)\bigr)P,& \text{if }\mathcal{S}_i(t)=D,
\end{cases}
\]
where
\[
x_i(t)=\frac{n_i(t)}{d_i}
\]
is the fraction of cooperative neighbors of \(i\). Thus payoffs are average payoffs per neighbor, not accumulated total payoffs.
This degree normalization keeps payoffs on the same scale across all vertices.

Aspirations are dynamic. After receiving payoff \(\pi_i(t)\),
individual \(i\) updates its aspiration recursively as a convex
combination of the previous aspiration and the current payoff:
\[
A_i(t+1)=(1-\rho)A_i(t)+\rho\pi_i(t),
\]
where \(0<\rho\le1\). Equivalently, we have
\[
A_i(t+1)-A_i(t)=\rho\bigl(\pi_i(t)-A_i(t)\bigr).
\]
The parameter \(\rho\) is the aspiration learning rate. Small \(\rho\) means that aspirations respond slowly to recent payoffs, while large \(\rho\) means that aspirations place more weight on the most recent payoff. The quantity \(r_i(t)=\pi_i(t)-A_i(t)\) is the payoff surplus relative to aspiration. A positive surplus means that the individual performed above aspiration, while a negative surplus represents dissatisfaction. Aspiration updating of this form is standard in adaptive aspiration-learning and satisficing models \citep{KarandikarMookherjeeRayVegaRedondo1998,MacyFlache2002,MasudaNakamura2011}. This formulation is also consistent with the concept of bounded rationality, in which individuals evaluate their realized payoffs against an acceptable performance threshold rather than seek a globally optimal strategy \citep{Simon1955}.

Strategy revision is aspiration-driven. Individual \(i\) switches
strategy with probability
\[
\sigma_i(t)=\frac{1}{1+\exp\{\beta[\pi_i(t)-A_i(t)]\}}.
\]
The parameter \(\beta\ge0\) is the intensity of selection. If
\(\beta=0\), switching has probability \(1/2\), independently of
payoff and aspiration. As \(\beta\) increases, dissatisfied individuals
become more likely to switch, while satisfied individuals become more
likely to keep their current strategy. Under weak selection, as \(\beta\to0^+\), we have
\[
\sigma_i(t)=\frac12-\frac{\beta}{4}\bigl(\pi_i(t)-A_i(t)\bigr)+O(\beta^3).
\]
The quadratic term vanishes by symmetry of the logistic function. The
revision rule is self-switching. A revising cooperator becomes a
defector, and a revising defector becomes a cooperator, so that for
every finite \(\beta\), both switching and not switching have strictly
positive probabilities.

\begin{figure}[H]
\centering
\includegraphics[width=\textwidth,height=20cm,keepaspectratio]{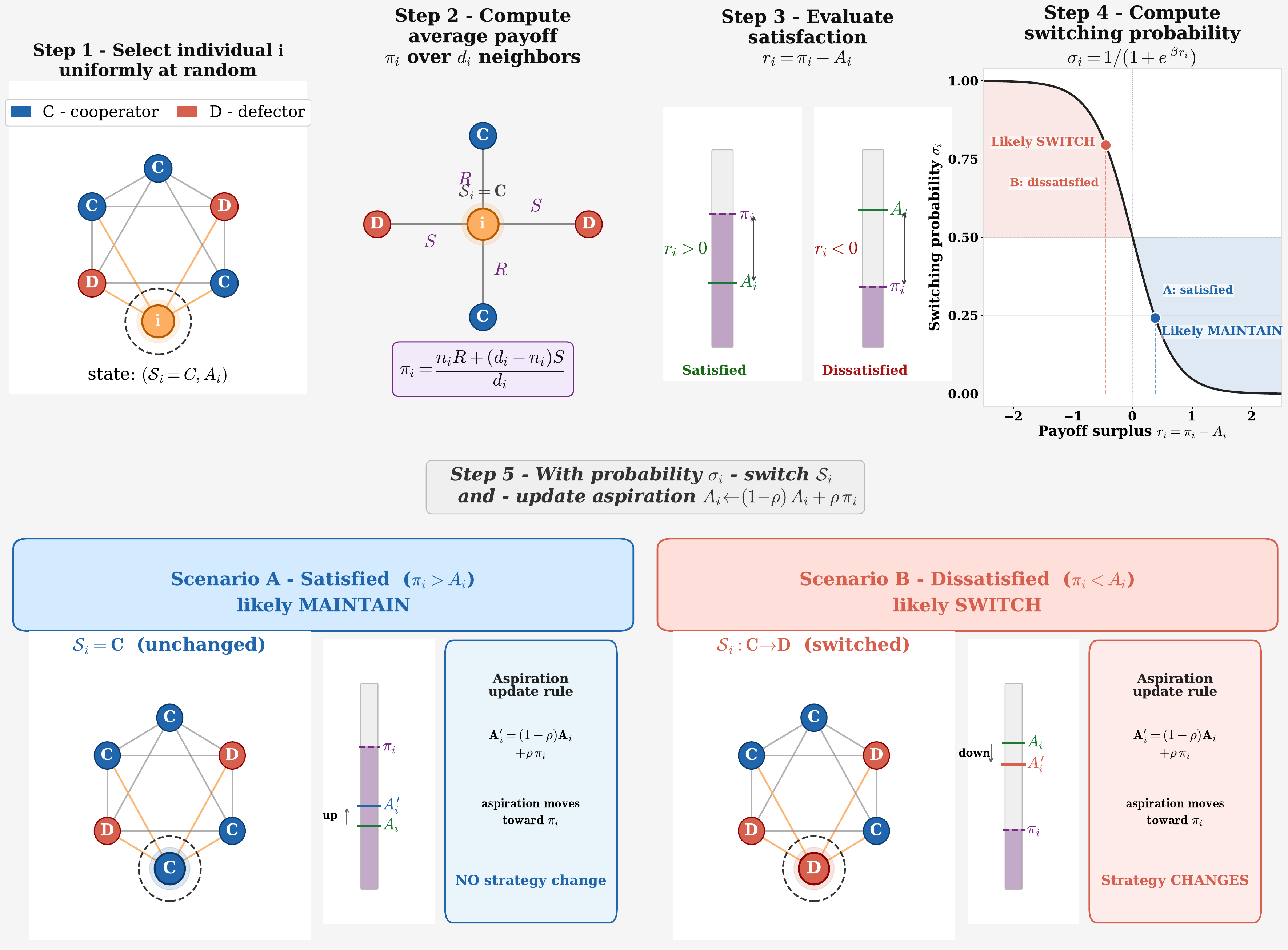}
\caption{
Schematic of one elementary update. First, an individual \(i\) is selected uniformly at random from the population. Its average payoff \(\pi_i\) is computed from pairwise interactions with its \(d_i\) neighbors. The payoff is then compared with the current aspiration level \(A_i\), giving the surplus \(r_i=\pi_i-A_i\). If \(r_i>0\), the individual is satisfied and the aspiration-based Fermi rule assigns a switching probability less than 1/2; if \(r_i<0\), the individual is dissatisfied and switching becomes more likely. Then, the individual either keeps its current strategy or self-switches to the opposite strategy with probability \(\sigma_i=(1+\exp\{\beta r_i\})^{-1}\). Finally, the aspiration level is updated as \(A_i'=(1-\rho)A_i+\rho\pi_i\). The lower panels illustrate the two possible outcomes. In both cases the updated aspiration is carried to the next time step. 
}
\label{fig:microscopic-update}
\end{figure}

One elementary update proceeds as follows. 1) Choose a vertex
\(I_t\in V\) uniformly at random. 2) Compute its current payoff
\(\pi_{I_t}(t)\) from its current strategy and neighborhood.
3) Compute the surplus
\(r_{I_t}(t)=\pi_{I_t}(t)-A_{I_t}(t)\). 4) With probability
\[
\sigma_{I_t}(t)=\frac{1}{1+\exp\{\beta[\pi_{I_t}(t)-A_{I_t}(t)]\}},
\]
the selected individual switches to the opposite strategy. Otherwise,
it keeps its current strategy. This means that
\[
\mathcal{S}_{I_t}(t+1)=
\begin{cases}
D,& \text{if }\mathcal{S}_{I_t}(t)=C \text{ and revision occurs},\\
C,& \text{if }\mathcal{S}_{I_t}(t)=D  \text{ and revision occurs},\\\mathcal{S}_{I_t}(t),
& \text{otherwise}.
\end{cases}
\]
For every \(j\neq I_t\), we set
\(\mathcal{S}_j(t+1)=\mathcal{S}_j(t)\). 5) Also, individual \(I_t\)
updates its aspiration level according to the rule
\begin{equation}\label{eq01}
A_{I_t}(t+1)=(1-\rho)A_{I_t}(t)+\rho\pi_{I_t}(t)=A_{I_t}(t)+\rho r_{I_t}(t),
\end{equation}
while all other aspirations remain unchanged.
Figure~\ref{fig:microscopic-update} summarizes the order of the
different steps in one elementary update.

The state of the population at time \(t\) is described by
\[
\boldsymbol{\mathcal Z}_t
=
\bigl(\boldsymbol{\mathcal S}(t),\mathbf A(t)\bigr),
\]
where $\boldsymbol{\mathcal S}(t)=\bigl(\mathcal S_i(t)\bigr)_{i\in V}\in\{C,D\}^N$ is the strategy profile and $\mathbf A(t)=\bigl(A_i(t)\bigr)_{i\in V}\in\mathbb R^N$ is the vector of aspiration levels. Thus, \((\boldsymbol{\mathcal Z}_t)_t\) is a discrete-time Markov process on $\Omega=\{C,D\}^N\times\mathbb R^N$.

This state space is not finite because the aspiration variables are
continuous. Define \(I=[\pi_-,\pi_+]\), where
\[
\pi_-=\min\{R,S,T,P\},
\qquad
\pi_+=\max\{R,S,T,P\}.
\]
For every vertex \(i\), \(\pi_i(t)\) is a convex combination of entries of the payoff matrix and belongs to \(I\), irrespective of the degree \(d_i\). Moreover, if \(A_i(t)\in I\), then
\[
A_i(t+1)=(1-\rho)A_i(t)+\rho\pi_i(t)\in I,
\]
because \(0<\rho\le1\) and both \(A_i(t)\) and \(\pi_i(t)\) lie in \(I\). Therefore,
\[
\mathcal E=\{C,D\}^N\times I^N
\]
is compact and forward invariant. Compactness follows from the finiteness of \(\{C,D\}^N\) and the compactness of \(I^N\) and forward invariance follows from the convexity of the aspiration update.

The initial aspirations are bounded but not necessarily in \(I\). Whenever a vertex is selected, its aspiration moves toward \(I\). Indeed, if \(M_i(t)\) denotes the number of times vertex \(i\) has been selected up to time \(t\) and 
$$\operatorname{dist}(x,I):=\inf_{a\in I}|x-a|,$$ then
\[
\operatorname{dist}(A_i(t),I)\le(1-\rho)^{M_i(t)}\operatorname{dist}(A_i(0),I).
\]
Since each vertex is selected infinitely often almost surely, every bounded initial aspiration vector is asymptotically attracted to \(\mathcal E\). Therefore, we study stationarity on the compact invariant space \(\mathcal E\).

The next result gives the stochastic basis for stationary abundance
used later. It concerns the process
\((\boldsymbol{\mathcal Z}_t)_t\) on the invariant state space
\(\mathcal E\).

\begin{theorem}\label{thm:stationary_distribution}
Let \(G=(V,E)\) be any finite connected graph with
\(N=|V|\ge2\), and suppose that each player's payoff is averaged over
its neighbors.
Then the process $(\boldsymbol{\mathcal{Z}}_t)_t$ restricted to $\mathcal E$ admits a unique invariant probability measure, denoted by \(\mu_N^\beta\). Moreover, for every bounded continuous function \(f:\mathcal E\to\mathbb R\) and $0<\beta\ll1$, we have
\[
\mathbb E_\beta[f]=\mathbb E_{0}[f]+o(1),
\]
where $\mathbb E_\beta$ and $\mathbb E_0$ denote the expectation with respect to the stationary distributions \(\mu_N^\beta\) and \(\mu_N^0\), respectively.
\end{theorem}

The proof is given in Appendix~\ref{appendixA}.

As a consequence, for any $0\le \beta<\infty$, the exact finite-population stationary abundance of
cooperation is well defined by
\[
\mathbb E_\beta\left[ p_C\right]=\int_{\mathcal E}p_C(\boldsymbol{\mathcal{S}})\,\mu_N^\beta(d\boldsymbol{\mathcal{S}},d\mathbf A).
\]

\begin{remark}
The assumptions \(0<\rho\le1\) and \(\beta<\infty\) are essential. If \(\rho=0\),
aspirations never adapt, and the continuous aspiration coordinates
can preserve memory of the initial condition. If \(\beta=\infty\),
some switching or non-switching probabilities may vanish, and
multiple invariant classes may appear.
\end{remark}

\begin{remark}
The measure \(\mu_N^\beta\) is the unique invariant
measure of the exact finite-population process. Its dependence on the
fixed graph \(G\) is suppressed in the notation.
The result above does not rely on a deterministic closure,
a large-population limit, a mean-field limit, or a
pair approximation.
\end{remark}

\begin{proposition}\label{lem:neutral_strategy_marginal}
At \(\beta=0\), the projection of \(\mu_N^0\) onto
the full strategy-profile space \(\{C,D\}^N\) is the uniform product
distribution. In particular, for every vertex \(i\), we have
\[
\mathbb P_{0}(\mathcal{S}_i=C)=\mathbb P_{0}(\mathcal{S}_i=D)=\frac12,
\]
and, conditional on either value of \(\mathcal{S}_i\), each neighbor of \(i\) is cooperative with probability \(1/2\).

\end{proposition}

\begin{proof}
For \(\beta=0\), the switching probability is
\(\sigma_i=\frac12\) for every vertex and every aspiration value.
Therefore, the strategy process is autonomous: at each elementary
update, a vertex is selected uniformly at random and its strategy is
flipped with probability \(1/2\), independently of payoffs,
aspirations, and neighboring strategies.

This finite Markov chain on \(\{C,D\}^N\) is irreducible, since any
strategy profile can be reached from any other by flipping the
vertices on which the two profiles differ. It is also aperiodic,
because at every state there is positive probability that the
selected vertex does not switch. Moreover, the transition probability
from \(\boldsymbol{\mathcal S}\) to
\(\boldsymbol{\mathcal S}'\) equals the transition probability from
\(\boldsymbol{\mathcal S}'\) to
\(\boldsymbol{\mathcal S}\): either
\(\boldsymbol{\mathcal S}=\boldsymbol{\mathcal S}'\), the two profiles
differ at exactly one vertex, or the transition probability is zero
in both directions. Hence the transition matrix is symmetric and
doubly stochastic.

The uniform distribution on \(\{C,D\}^N\) is therefore stationary. By
irreducibility, it is the unique stationary distribution of the
neutral strategy chain. Since the joint invariant measure
\(\mu_N^0\) projects onto an invariant measure for the chain on the full strategy-profile space
\(\{C,D\}^N\), this projection must be the uniform product distribution.

Under this distribution, the indicators
\((\mathbf{1}_{\{\mathcal S_j=C\}})_{j\in V}\) are independent Bernoulli
variables with parameter \(1/2\). Therefore, conditional on
\(\mathcal S_i=C\) or \(\mathcal S_i=D\), each neighbor of \(i\) is
cooperative with probability \(1/2\).
\end{proof}

\begin{remark}
Under the neutral stationary distribution, strategies
at adjacent vertices are independent, so there is no assortment along
edges. This follows from the self-switching rule: at \(\beta=0\),
strategy revision is independent of neighboring strategies.  In contrast, neutral
social-learning or inheritance models can generate correlations when
an individual copies the strategy of a neighbor or parent.
\end{remark}

\section{\label{sec:exact-drift}Exact Drift Identities and Weak-Selection Abundance}
This section is organized as follows. We first derive the exact one-step drift of the cooperator frequency and then expand this drift for weak selection. Next, we compute the aspiration-sum identities needed to separate payoff effects from aspiration effects at stationarity. These identities are then combined to obtain the first-order stationary abundance formula. Finally, we record the neutral variance and state the first-order variance correction, which shows how weak selection changes the fluctuations of the cooperator frequency around its stationary mean.

Throughout, we condition on the full 
state \(\boldsymbol{\mathcal{Z}}_t=(\boldsymbol{\mathcal{S}}(t),\mathbf A(t))\). To simplify notation, we suppress the
time variable.

\subsection{Cooperator frequency drifts}
First, we compute the one-step drift of the cooperator frequency.
Since only one individual is selected in an elementary update, the
number of cooperators, denoted by \(X\), can change
only by \(+1\), \(-1\), or \(0\). A switching defector increases \(X\)
by one, while a switching cooperator decreases \(X\) by one. Therefore,
if
\[
\Delta X=X(t+1)-X(t),
\]
then, we have
\begin{subequations}
\begin{align}
\mathbb P_\beta \bigl(\Delta X=+1\mid\boldsymbol{\mathcal S},\mathbf A\bigr)&=\frac1N\sum_{i:\mathcal S_i=D}\sigma_i,\\
\mathbb P_\beta \bigl(\Delta X=-1\mid\boldsymbol{\mathcal S},\mathbf A\bigr)&=\frac1N\sum_{i:\mathcal S_i=C}\sigma_i,\\
\mathbb P_\beta \bigl(\Delta X=0\mid\boldsymbol{\mathcal S},\mathbf A\bigr)&=1-\frac1N\sum_{i\in V}\sigma_i.
\end{align}
\end{subequations}
Since \(p_C=X/N\), the one-step change in the frequency of cooperation
is
\[
\Delta p_C=p_C(t+1)-p_C(t)=\frac{\Delta X}{N}.
\]
Consequently, the conditional expected change in the frequency of
cooperation can be written as
\begin{equation}\label{eq50}
\mathbb E_\beta\!\left[ \Delta p_C\mid\boldsymbol{\mathcal S},\mathbf A\right]=\frac1{N^2}\left( \sum_{i:\mathcal S_i=D}\sigma_i - \sum_{i:\mathcal S_i=C}\sigma_i\right).
\end{equation}

Under weak selection, as \(\beta\to0^+\), the Fermi
function satisfies
\[
\sigma_i=\frac12-\frac{\beta}{4}r_i+O(\beta^3).
\]
Since the process is restricted to the compact state space
\(\mathcal E\), the \(O(\beta^3)\) remainder is uniform on
\(\mathcal E\). Moreover, degree normalization makes
this bound independent of the degree \(d_i\). Substituting this expansion into
Eq.~\eqref{eq50} gives
\begin{equation}\label{eq4}
\mathbb E_\beta\!\left[ \Delta p_C\mid\boldsymbol{\mathcal S},\mathbf A\right]=\frac{1-2p_C}{2N}+\frac{\beta}{4N^2}\left[ \sum_{i:\mathcal S_i=C}r_i - \sum_{i:\mathcal S_i=D}r_i\right]+O\!\left(\beta^3\right).
\end{equation}
\subsection{Aspiration-sum drifts}
We also need the drift of the total aspiration carried by each strategy class. For that, define $M_C=\sum_{i:\mathcal{S}_i=C}A_i$ and $M_D=\sum_{i:\mathcal{S}_i=D}A_i$. 
Their one-step increments are denoted by
\[
\begin{aligned}
\Delta M_C&=M_C(t+1)-M_C(t),\\
\Delta M_D&=M_D(t+1)-M_D(t).
\end{aligned}
\]
First consider a selected cooperator \(i\). If it does not switch,
then it remains in the cooperator class and its aspiration changes
from \(A_i(t)\) to
\(
A_i(t+1)=A_i(t)+\rho r_i(t).
\)
Hence, the changes in the total aspirations of the two strategy classes are
\[
\begin{aligned}
\Delta M_C
&=A_i(t+1)-A_i(t)=\rho r_i(t),\\
\Delta M_D&=0.
\end{aligned}
\]
If it switches from \(C\) to \(D\), then the cooperator class loses
the old aspiration \(A_i(t)\), while the defector class gains the
updated aspiration \(A_i(t+1)\), leading to
\[
\begin{aligned}
\Delta M_C&=-A_i(t),\\
\Delta M_D&=A_i(t+1).
\end{aligned}
\]

By symmetry, if the selected individual \(i\) is a defector and does
not switch, then
\[
\begin{aligned}
\Delta M_C&=0,\\
\Delta M_D&=A_i(t+1)-A_i(t)=\rho r_i(t).
\end{aligned}
\]
If it switches from \(D\) to \(C\), then
\[
\begin{aligned}
\Delta M_C&=A_i(t+1),\\
\Delta M_D&=-A_i(t).
\end{aligned}
\]

Averaging over the uniformly selected individual and the
switch/no-switch event gives the exact conditional drifts
\begin{equation}\label{eq10}
\begin{aligned}
\mathbb E_\beta\!\left[ \Delta M_C\mid\boldsymbol{\mathcal S},\mathbf A\right]&=\frac1N\Bigg[ \sum_{i:\mathcal S_i=C} \left(   (1-\sigma_i)\rho r_i(t)-\sigma_iA_i(t) \right) + \sum_{i:\mathcal S_i=D} \sigma_i \bigl(A_i(t)+\rho r_i(t)\bigr)\Bigg],\\
\mathbb E_\beta\!\left[\Delta M_D\mid\boldsymbol{\mathcal S},\mathbf A\right]&=\frac1N\Bigg[ \sum_{i:\mathcal S_i=D} \left(   (1-\sigma_i)\rho r_i(t)-\sigma_iA_i(t) \right) + \sum_{i:\mathcal S_i=C} \sigma_i\bigl(A_i(t)+\rho r_i(t)\bigr)\Bigg].
\end{aligned}
\end{equation}

Define also the total aspiration level
\[
M=M_C+M_D=\sum_{i=1}^N A_i
\]
and its increment
\[
\Delta M=M(t+1)-M(t).
\]
Adding the two identities in \eqref{eq10} gives the drift of \(M\).
Indeed, the switch/no-switch event only determines whether the updated
aspiration of the selected individual is counted in \(M_C\) or in
\(M_D\). It does not create any additional change in the total
aspiration level. The only contribution to \(\Delta M\) comes from
the aspiration update at the selected vertex:
\[
A_i(t+1)-A_i(t)=\rho r_i(t).
\]
Therefore,
\begin{equation}\label{eq20}
\mathbb E_\beta\!\left[ \Delta M\mid\boldsymbol{\mathcal S},\mathbf A\right]=\frac{\rho}{N}\sum_{i\in V}r_i.
\end{equation}
In particular, this conditional drift has no explicit
dependence on the vertex degrees or on the selection intensity
\(\beta\).
\subsection{\label{subsec:exact-first-order-abundance}
Exact first-order stationary abundance}

Now, we will use the drift identity for \(p_C\) to derive
the leading weak-selection formula for the stationary abundance of
cooperation, \(\mathbb E_\beta[p_C]\). The result is exact for the
finite stochastic process and does not rely on a pair approximation,
deterministic closure, or a large-population limit.

\begin{theorem}\label{thm:exact_first_order_abundance}
Let \(G=(V,E)\) be any finite connected graph with
\(2\le N=|V|<\infty\). Fix \(0<\rho\le1\). For $0<\beta\ll1$, we have
\[
\mathbb E_\beta[p_C]=\frac12+\frac{\beta}{16}\left(R+S-T-P\right)+o(\beta).
\]
\end{theorem}

\begin{proof}
First, note the stationarity identity that will be used below. If
\(F:\mathcal E\to\mathbb R\) is bounded and
$\boldsymbol{\mathcal Z}_t=\bigl(\boldsymbol{\mathcal S}(t),\mathbf A(t)\bigr)$
is stationary with law \(\mu_N^\beta\), then
\[
\mathbb E_\beta\!\left[ F(\boldsymbol{\mathcal Z}_{t+1}) - F(\boldsymbol{\mathcal Z}_t)\right]=0.
\]
Applying this identity to Eq. \eqref{eq4}, after taking expectations and using the tower property, we obtain
\[
0=\mathbb E_\beta\left[\Delta p_C\right]=\mathbb E_\beta\left[\frac{1-2p_C}{2}+\frac{\beta}{4N}\left(\sum_{i:\mathcal{S}_i=C}r_i-\sum_{i:\mathcal{S}_i=D}r_i\right)\right]+o(\beta).
\]
Hence
\[
\mathbb E_\beta\left[ p_C\right]=\frac12+\frac{\beta}{4N}\mathbb E_\beta\left[\sum_{i:\mathcal{S}_i=C}r_i-\sum_{i:\mathcal{S}_i=D}r_i\right]+o(\beta).
\]
On the other hand, by Theorem~\ref{thm:stationary_distribution}, it follows that
\[
\mathbb E_\beta[\cdot]=\mathbb E_{0}[\cdot]+o(1).
\]
Therefore, we have
\begin{equation}\label{eq13}
\mathbb E_\beta\left[ p_C\right]=\frac12+\frac{\beta}{4N}\mathbb E_{0}\left[\sum_{i:\mathcal{S}_i=C}r_i-\sum_{i:\mathcal{S}_i=D}r_i\right]+o(\beta).
\end{equation}

It remains to compute \(\mathbb E_{0}\left[\sum_{i:\mathcal{S}_i=C}r_i-\sum_{i:\mathcal{S}_i=D}r_i\right]\). By Proposition~\ref{lem:neutral_strategy_marginal}, the strategy marginal of \(\mu_N^0\) is uniform on \(\{C,D\}^N\). Hence
\[
\mathbb P_{0}(\mathcal{S}_i=C)=\mathbb P_{0}(\mathcal{S}_i=D)=\frac12,
\]
and, conditional on either value of \(\mathcal{S}_i\), each neighbor of \(i\) is cooperative with probability \(1/2\).

Since \(d_i\ge1\), for either \(s\in\{C,D\}\), we have
\[
\begin{aligned}
\mathbb E_0[x_i\mid\mathcal S_i=s]&=\frac1{d_i}\sum_{j\sim i}\mathbb P_0(\mathcal S_j=C\mid\mathcal S_i=s)=\frac1{d_i}\sum_{j\sim i}\frac12=\frac12.
\end{aligned}
\]
Here, $x_i$ is the fraction of cooperative neighbors of individual $i$.
Consequently, the expected payoff is
\[
\begin{aligned}
&\mathbb E_0[\pi_i\mid\mathcal S_i=C]=\frac{R+S}{2},\\
&\mathbb E_0[\pi_i\mid\mathcal S_i=D]=\frac{T+P}{2}.
\end{aligned}
\]
In particular, these conditional expectations do not depend on \(d_i\). Therefore, we obtain
\begin{equation}\label{eq11}
\begin{aligned}
\mathbb E_0\!\left[\sum_{i:\mathcal S_i=C}\pi_i\right]&=\frac{N}{2}\Pi_C^0,\\
\mathbb E_0\!\left[\sum_{i:\mathcal S_i=D}\pi_i\right]&=\frac{N}{2}\Pi_D^0,
\end{aligned}
\end{equation}
where
\[
\Pi_C^0=\frac{R+S}{2},
\qquad
\Pi_D^0=\frac{T+P}{2}.
\]

We next compute the aspiration term. At neutrality,
\(\sigma_i=1/2\). Using
\[
A_i(t+1)=A_i(t)+\rho r_i(t)
\]
in Eq.~\eqref{eq10}, the contribution of a current cooperator \(i\) to the drift of \(M_C-M_D\) is
\[
\frac12\rho r_i-\frac12A_i-\frac12\bigl(A_i+\rho r_i\bigr)=-A_i,
\]
whereas the contribution of a current defector \(i\) is
\[
\frac12\bigl(A_i+\rho r_i\bigr)-\left(\frac12\rho r_i-\frac12A_i\right)=A_i.
\]
It follows that
\begin{equation}\label{eq14}
\mathbb E_0\!\left[ \Delta(M_C-M_D)\mid\boldsymbol{\mathcal S},\mathbf A\right]=\frac1N(M_D-M_C).
\end{equation}
Taking expectation with respect to \(\mu_N^0\) and using stationarity,
we obtain
\[
0=\mathbb E_0[\Delta(M_C-M_D)]=\frac1N\mathbb E_0[M_D-M_C].
\]
Hence
\begin{equation}\label{eq12}
\mathbb E_0\!\left[ \sum_{i:\mathcal S_i=D}A_i - \sum_{i:\mathcal S_i=C}A_i\right]=0.
\end{equation}

Combining Eqs \eqref{eq11} and \eqref{eq12}, we obtain
\[
\begin{aligned}
\mathbb E_{0}\left[\sum_{i:\mathcal{S}_i=C}r_i-\sum_{i:\mathcal{S}_i=D}r_i\right]&=\mathbb E_{0}\left[\sum_{i:\mathcal{S}_i=C}(\pi_i-A_i)-\sum_{i:\mathcal{S}_i=D}(\pi_i-A_i)\right]\\
&=\mathbb E_{0}\left[\sum_{i:\mathcal{S}_i=C}\pi_i-\sum_{i:\mathcal{S}_i=D}\pi_i\right]-\mathbb E_{0}\left[\sum_{i:\mathcal{S}_i=C}A_i-\sum_{i:\mathcal{S}_i=D}A_i\right]\\
&=\frac{N}{2}(\Pi_C^0-\Pi_D^0).
\end{aligned}
\]
Substituting the last identity into the stationary drift identity \eqref{eq13} completes the proof.
\end{proof}

\begin{remark}
By Theorem~\ref{thm:exact_first_order_abundance}, for all sufficiently small $\beta > 0$, the stationary abundance of cooperation strictly exceeds its neutral value if $R+S > T+P$, and falls strictly below it if $R+S < T+P$. When $R+S = T+P$, the first-order correction vanishes, providing no information about the direction of higher-order shifts.
\end{remark}

\begin{remark}\label{rem:graph-independence}
Theorem~\ref{thm:exact_first_order_abundance} gives an exact
first-order weak-selection result for the finite process,
independent of \(N\), the degree sequence, and the
topology of the underlying graph. It is not a closure prediction and
does not require a large-population limit.

This graph independence has two sources. First, the
self-switching rule does not involve social learning: no strategy is
copied from a neighbor, and no individual competes to reproduce so the
usual effects of assortment and reproductive competition are absent.
Second, averaging payoffs by the focal vertex's degree removes the
remaining degree scale from the first-order calculation.

For accumulated payoffs, a \(k\)-regular graph produces a factor \(k\)
in the first-order coefficient, which is equivalent at this order to
rescaling the selection intensity. On an irregular graph, the
accumulated payoff scale is vertex dependent, so the full
finite-selection process cannot be obtained from the averaged-payoff
process through a single global rescaling of \(\beta\).
\end{remark}

\subsection{\label{ssec:neutral-fluctuations}Neutral fluctuations and first-order variance}
In this section, we use the drift identities to derive two main results. First, we obtain the stationary variance of cooperator abundance under neutrality. Second, we calculate the first-order weak-selection expansion of the stationary variance. These results clarify how weak selection affects the statistical fluctuations around the stationary mean.
\subsubsection{Stationary variance under neutrality}

\begin{proposition}\label{prop:neutral-variance}
For every finite connected graph \(G=(V,E)\), the stationary variance at neutrality is
\[
\operatorname{Var}_{0}(p_C)=\frac{1}{4N}.
\]
\end{proposition}
\begin{proof}
The neutral strategy marginal of \(\mu_N^0\) is the uniform product measure on \(\{C,D\}^N\), where the random variables $\left(\mathbf 1_{\{\mathcal{S}_i=C\}}\right)_{i\in V}$ are independent Bernoulli random variables with parameter \(1/2\). Thus $p_C=\frac1N\sum_{i\in V}\mathbf 1_{\{\mathcal{S}_i=C\}}$ has the same distribution as $\frac1N\operatorname{Binomial}\left(N,\frac12\right)$. Consequently, we obtain
\[
\operatorname{Var}_{0}(p_C)= \frac{1}{N^2}\,N\,\frac12\,\left(1-\frac12\right) = \frac{1}{4N}.
\]
\end{proof}
\subsubsection{First-order correction under weak selection}

\begin{theorem}\label{thm:variance-first-order}

For every finite connected graph \(G=(V,E)\) and for \(0<\beta\ll1\), we have
\[
\operatorname{Var}_{\beta}(p_C)=\frac{1}{4N}+\frac{\beta}{32N}\bigl[(R-S)-(T-P)\bigr]+o(\beta).
\]
\end{theorem}

In the proof, we use the following shorthands
\[
\begin{aligned}
Q^{(A)}&=\sum_{i:\mathcal{S}_i=C}A_i-\sum_{i:\mathcal{S}_i=D}A_i,\\
Q^{(\pi)}&=\sum_{i:\mathcal{S}_i=C}\pi_i-\sum_{i:\mathcal{S}_i=D}\pi_i,\\
Q&=\sum_{i:\mathcal{S}_i=C}r_i-\sum_{i:\mathcal{S}_i=D}r_i,
\end{aligned}
\]
so that $Q=Q^{(\pi)}-Q^{(A)}$.

The proof is organized as follows. Lemma~\ref{lem:pQ-neutral} computes the neutral contribution coming from the aspiration balance \(Q^{(A)}\), while Lemma~\ref{lem:pHpi-neutral} computes the corresponding neutral contribution coming from the payoff balance \(Q^{(\pi)}\). These two identities are then combined in Lemma~\ref{lem:second-moment-expansion} to obtain the weak-selection expansion of \(\mathbb E_\beta[p_C^2]\). Finally, the proof of Theorem~\ref{thm:variance-first-order} subtracts the square of the first-order abundance expansion given in Theorem~\ref{thm:exact_first_order_abundance} from this second-moment expansion to obtain the stationary variance formula.
\begin{lemma}\label{lem:pQ-neutral}
Under neutrality, we have
\[
\mathbb E_{0}\left[p_CQ^{(A)}\right]=\frac{R+S+T+P}{8}.
\]
\end{lemma}

\begin{proof}
 Since
\begin{equation}\label{eq51}
\Delta(p_CQ^{(A)})=Q^{(A)}\Delta p_C+p_C\Delta Q^{(A)}+\Delta p_C\Delta Q^{(A)},
\end{equation}
we need to compute the three conditional contributions separately.

On the one hand, evaluating the drift identity \eqref{eq4} at \(\beta=0\) gives
\begin{equation}\label{eq52}
\mathbb E_0[\Delta p_C\mid \boldsymbol{\mathcal{S}},\mathbf A]=\frac{1-2p_C}{2N}.
\end{equation}
On the other hand, Eq.~\eqref{eq14}, written in terms of \(Q^{(A)}=M_C-M_D\), gives
\begin{equation}\label{eq53}
\mathbb E_0[\Delta Q^{(A)}\mid \boldsymbol{\mathcal{S}},\mathbf A]=-\frac{Q^{(A)}}{N}.
\end{equation}
It remains to compute the mixed term. If a cooperator \(i\) switches, then $\Delta p_C=-1/N$ and $\Delta Q^{(A)}=-2A_i-\rho r_i$. If a defector \(i\) switches, then $\Delta p_C=1/N$ and $\Delta Q^{(A)}=2A_i+\rho r_i$. In both cases, we obtain
\[
\Delta p_C\Delta Q^{(A)}=\frac{2A_i+\rho r_i}{N}.
\]
At neutrality, the selected individual switches with probability \(1/2\). Hence, we have
\begin{equation}\label{eq54}
\mathbb E_0\left[\Delta p_C\Delta Q^{(A)}\mid \boldsymbol{\mathcal{S}},\mathbf A\right]=\frac{M}{N^2}+\frac{\rho}{2N^2}\sum_{i\in V}r_i.
\end{equation}
Combining the three terms in Eqs. \eqref{eq52}-\eqref{eq54} in Eq. \eqref{eq51} after taking its conditional expectation yields
\[
\mathbb E_0\left[\Delta(p_CQ^{(A)})\mid \boldsymbol{\mathcal{S}},\mathbf A\right]=\frac{Q^{(A)}}{2N}-\frac{2p_CQ^{(A)}}{N}+\frac{M}{N^2}+\frac{\rho}{2N^2}\sum_{i\in V}r_i.
\]

Taking expectation under \(\mu_N^0\) and using stationarity gives
\[
0=\mathbb E_{0}\left[\Delta(p_CQ^{(A)})\right]
=\frac{1}{2N}\mathbb E_{0}\left[Q^{(A)}\right]-\frac{2}{N}\mathbb E_{0}\left[p_CQ^{(A)}\right]+\frac{1}{N^2}\mathbb E_{0}[M]+\frac{\rho}{2N^2}\mathbb E_{0}\left[\sum_{i\in V}r_i\right],
\]
from which we deduce that
\begin{equation}\label{eq30}
\mathbb E_{0}\left[p_CQ^{(A)}\right]=\frac14\mathbb E_{0}\left[Q^{(A)}\right]+\frac{\rho}{4N}\mathbb E_{0}\left[\sum_{i\in V}r_i\right]+\frac{1}{2N}\mathbb E_{0}[M].
\end{equation}

Now, we will compute the three terms in the last identity. The first term is given by Eq.~\eqref{eq12},
\[
\mathbb E_{0}\left[Q^{(A)}\right]=0. 
\]
For the second term, using stationarity of the total aspiration \(M\), together with Eq.~\eqref{eq20} for $\beta=0$, we get
\begin{equation}\label{eq35}
\mathbb E_{0}\left[\sum_{i\in V}r_i\right]=0.
\end{equation}
The third term is a consequence of Eq.~\eqref{eq11} as
\[
\mathbb E_{0}[M]=\mathbb E_{0}\left[\sum_{i\in V}A_i\right]=\mathbb E_{0}\left[\sum_{i\in V}\pi_i\right]-\mathbb E_{0}\left[\sum_{i\in V}r_i\right]=\frac{N}{2}(\Pi_C^0+\Pi_D^0).
\]
Substituting these identities into Eq. \eqref{eq30}, the result follows.
\end{proof}


\begin{lemma}\label{lem:pHpi-neutral}
Under neutrality, we have
\[
\mathbb E_0\left[p_CQ^{(\pi)}\right]=\frac{(N+2)R+NS-NT-(N-2)P}{8}.
\]
\end{lemma}

\begin{proof}
Note that $Q^{(\pi)}$ can be written as 
$Q^{(\pi)}=\sum_{i\in V}\epsilon_i\pi_i$, where $\epsilon_i=\mathbf 1_{\{\mathcal{S}_i=C\}}-\mathbf 1_{\{\mathcal{S}_i=D\}}$. Let $X=\sum_{j\in V}\mathbf 1_{\{\mathcal{S}_j=C\}}$ denote the number of cooperators in the population. Then, we have
\begin{equation}\label{eq31}
\mathbb E_0\left[p_CQ^{(\pi)}\right]=\frac1N\sum_{i\in V}\mathbb E_0\left[X\epsilon_i\pi_i\right].
\end{equation}

Fix \(i\in V\), and let \(d_i\) be its degree. We decompose \(X\) as
\[
X=\mathbf 1_{\{\mathcal S_i=C\}}+\sum_{j\sim i}\mathbf 1_{\{\mathcal S_j=C\}}+\sum_{\substack{j\neq i\\j\not\sim i}}\mathbf 1_{\{\mathcal S_j=C\}}.
\]

For the contribution of vertex \(i\) itself, we have
\begin{equation}\label{eq32}
\mathbb E_0\left[ \mathbf 1_{\{\mathcal S_i=C\}}\epsilon_i\pi_i\right]=\frac12\Pi_C^0=\frac{R+S}{4}.
\end{equation}

Next, fix a neighbor \(j\sim i\). Conditional on
\(\mathcal S_j=C\), the remaining \(d_i-1\) neighbors of \(i\) are
independent cooperators with probability \(1/2\). Therefore,
\[
\mathbb E_0\left[\pi_i\mid\mathcal S_i=C,\mathcal S_j=C\right]=\frac{R+(d_i-1)\Pi_C^0}{d_i}=\frac{(d_i+1)R+(d_i-1)S}{2d_i},
\]
and
\[
\mathbb E_0\left[\pi_i\mid\mathcal S_i=D,\mathcal S_j=C\right]=\frac{T+(d_i-1)\Pi_D^0}{d_i}=\frac{(d_i+1)T+(d_i-1)P}{2d_i}.
\]
Multiplying these two cases by their neutral probabilities \(1/4\), and by the
signs \(+1\) and \(-1\) coming from \(\epsilon_i\), gives 
\[
\mathbb E_0\left[ \mathbf 1_{\{\mathcal S_j=C\}}\epsilon_i\pi_i\right]=\frac{ (d_i+1)R+(d_i-1)S -(d_i+1)T-(d_i-1)P}{8d_i}.
\]
Summing over the \(d_i\) neighbors of \(i\), we obtain 
\begin{equation}\label{eq33}
\sum_{j\sim i}\mathbb E_0\left[\mathbf 1_{\{\mathcal S_j=C\}}\epsilon_i\pi_i\right]=\frac{ (d_i+1)R+(d_i-1)S -(d_i+1)T-(d_i-1)P}{8}.
\end{equation}

For a non-neighbor $j$ (\(j\neq i\) and \(j\not\sim i\)), the variable \(\mathcal{S}_j\) is independent of \((\mathcal{S}_i,\pi_i)\) under the neutral strategy marginal. Hence, we have
\[
\begin{aligned}
\mathbb E_0\left[\mathbf 1_{\{\mathcal S_j=C\}}\epsilon_i\pi_i\right]=\frac12\mathbb E_0[\epsilon_i\pi_i]=\frac14\left(\Pi_C^0-\Pi_D^0\right)=\frac{R+S-T-P}{8}.
\end{aligned}
\]
There are \(N-1-d_i\) such vertices, so their total contribution is 
\begin{equation}\label{eq34}
\sum_{\substack{j\neq i\\j\not\sim i}}\mathbb E_0\left[ \mathbf 1_{\{\mathcal S_j=C\}}\epsilon_i\pi_i\right]=\frac{(N-1-d_i)(R+S-T-P)}{8}.
\end{equation}

Combining Eqs.~\eqref{eq32}--\eqref{eq34}, we find
\[
\mathbb E_0[X\epsilon_i\pi_i]=\frac{(N+2)R+NS-NT-(N-2)P}{8}.
\]
The degree \(d_i\) has canceled, so the last expression is the same
for every vertex \(i\), even though the graph need not be regular or
vertex-transitive. Substitution into Eq.~\eqref{eq31} gives
\[
\mathbb E_0\left[p_CQ^{(\pi)}\right]=\frac{(N+2)R+NS-NT-(N-2)P}{8},
\]
which proves the lemma.
\end{proof}


\begin{lemma}\label{lem:second-moment-expansion}
We have
\begin{equation}\label{eq40}
\mathbb E_\beta[p_C^2]=\frac14+\frac{1}{4N}+\frac{\beta}{32N}\left[(2N+1)R+(2N-1)S-(2N+1)T-(2N-1)P\right]+o(\beta).
\end{equation}
\end{lemma}

\begin{proof}
First, we will compute the drift of \(p_C^2\). Since
\[
\Delta(p_C^2)=2p_C\Delta p_C+(\Delta p_C)^2,
\]
there are two contributions to consider.

On the one hand, the drift identity \eqref{eq4} gives
\[
\mathbb E_\beta[\Delta p_C\mid \boldsymbol{\mathcal{S}},\mathbf A]=\frac{1-2p_C}{2N}+\frac{\beta Q}{4N^2}+o(\beta).
\]
On the other hand, we have \((\Delta p_C)^2=1/N^{2}\)  when the selected individual switches. Otherwise, it vanishes. Therefore, we obtain
\[
\mathbb E_\beta[(\Delta p_C)^2\mid \boldsymbol{\mathcal{S}},\mathbf A]=\frac1{N^3}\sum_{i\in V}\sigma_i=\frac{1}{2N^2}-\frac{\beta}{4N^3}\sum_{i\in V}r_i+o(\beta).
\]
Putting these two estimates together yields
\[
\mathbb E_\beta[\Delta(p_C^2)\mid \boldsymbol{\mathcal{S}},\mathbf A]=\frac{p_C(1-2p_C)}{N}+\frac{1}{2N^2}+\beta\left(\frac{p_CQ}{2N^2}-\frac{1}{4N^3}\sum_{i\in V}r_i\right)+o(\beta).
\]

Now, taking expectation with respect to \(\mu_N^\beta\) and using the stationarity identity $\mathbb E_\beta[\Delta(p_C^2)]=0$, we obtain
\begin{equation}\label{eq36}
\mathbb E_\beta[p_C^2]=\frac{1}{2}\mathbb E_\beta[p_C]+\frac{1}{4N}+\beta\left(\frac{1}{4N}\mathbb E_\beta\left[p_CQ\right]-\frac{1}{8N^2}\mathbb E_\beta\left[\sum_{i\in V}r_i\right]\right)+o(\beta).
\end{equation} 
For the first term, it is given in Theorem \ref{thm:exact_first_order_abundance} by
\begin{equation}\label{eq37}
\mathbb E_\beta\left[ p_C\right]=\frac12+\beta\frac{R+S-T-P}{16}+o(\beta).
\end{equation} 
For the second term, using Lemmas \ref{lem:pQ-neutral} and \ref{lem:pHpi-neutral} and the fact that $Q=Q^{(\pi)}-Q^{(A)}$, we obtain
\begin{equation}\label{eq38}
\begin{aligned}
\mathbb E_\beta\left[p_CQ\right]&=\mathbb E_0\left[p_CQ\right]+o(1)\\
&=\mathbb E_0\left[p_CQ^{(\pi)}\right]-\mathbb E_0\left[p_CQ^{(A)}\right]+o(1)\\
&=\frac{(N+1)R+(N-1)S-(N+1)T-(N-1)P}{8}+o(1).
\end{aligned}
\end{equation}
For the last term,  using stationarity of the total aspiration \(M\), together with Eq.~\eqref{eq20}, we have
\begin{equation}\label{eq39}
\mathbb E_{\beta}\left[\sum_{i\in V}r_i\right]=0.
\end{equation} 
Inserting Eqs. \eqref{eq37}--\eqref{eq39} into \eqref{eq36} gives the expression in \eqref{eq40}.
\end{proof}


\begin{proof}[Proof of Theorem \ref{thm:variance-first-order}]
Using Lemma~\ref{lem:second-moment-expansion} and Theorem~\ref{thm:exact_first_order_abundance}, we get
\[
\begin{aligned}
\operatorname{Var}_{\beta}(p_C)&=\mathbb E_\beta\left[p_C^2\right]-\left(\mathbb E_\beta[p_C]\right)^2\\
&= \frac14+\frac{1}{4N}+\frac{\beta}{32N}\left[(2N+1)R+(2N-1)S-(2N+1)T-(2N-1)P\right]\\
&\qquad- \left(\frac14+\frac{\beta}{16}(R+S-T-P)\right)+o(\beta)\\
&=\frac{1}{4N}+\frac{\beta}{32N}\bigl[R-S-T+P\bigr]+o(\beta).
\end{aligned}
\]
This gives the announced expansion.
\end{proof}

\begin{remark}
The variance correction has a different interpretation from the
abundance correction. While the mean abundance is controlled by the
average payoff contrast
\[
R+S-T-P,
\]
the variance is controlled by
\[
(R-S)-(T-P).
\]
Here, \(R-S\) is the difference between the payoff received by a
cooperator when all its neighbors are cooperators and its payoff when
all its neighbors are defectors. Thus, it is the amount
by which the cooperator's payoff changes when an all-cooperative
neighborhood becomes all defecting. Similarly, \(T-P\) is the
corresponding payoff difference for a defector. Therefore, the
first-order variance increases when \(R-S>T-P\), decreases when
\(R-S<T-P\), and has a zero first-order correction when
\(R-S=T-P\). For fixed \(N\), this first-order variance correction
is independent of the topology of the finite
connected graph.
\end{remark}

\begin{remark}
The preceding variance formula is a first-order weak-selection result. Higher-order corrections are more delicate. Although the pointwise Taylor expansion of the Fermi switching probability contains only odd powers of \(\beta\), this does not by itself imply that stationary quantities have only odd powers of \(\beta\). The invariant measure \(\mu_N^\beta\) also depends on \(\beta\), and its perturbation can contribute additional terms. In particular, computing second-order corrections would require control of the perturbation of the invariant measure beyond first order. Such calculations involve mixed moments of current strategies, current payoffs, and endogenous aspirations.
\end{remark}


\section{\label{sec:applications}Applications to Standard Pairwise Games}

In this section, we apply our analytical framework to four standard pairwise games: the donation game, the snowdrift game, the stag-hunt game, and harmony games. We examine how weak selection shifts both the expected abundance of cooperation and its dispersion around the mean.


\subsection{Donation game}
For the donation game, the payoff vector is $(R,S,T,P)=(b-c,-c,b,0)$, where $b>c>0$. Since
\[
R+S-T-P=-2c<0,
\]
weak selection favors the abundance of defection. Moreover, as
\[
R-S=T-P=b,
\]
the first-order variance correction is zero. Consequently, weak selection shifts the stationary mean against cooperation in the donation game, but leaves the stationary variance unchanged at first order.

Biologically, this reflects the cost structure of donation. For any fixed local environment, a cooperator receives exactly $c$ less than a defector facing the same neighborhood: cooperation provides benefits to others but pays the cost of helping. This makes cooperators more likely, on average, to fall below their aspiration and revise their strategy. Defectors, by avoiding the cost, are more likely to be satisfied. This explains why the stationary abundance of cooperation is shifted below its neutral value.

The zero first-order variance correction has a direct interpretation. 
A cooperator's payoff decreases by \(R-S=b\) when its neighborhood
changes from all cooperative to all defecting. The corresponding
decrease for a defector is also \(T-P=b\). Thus, the payoffs of
cooperators and defectors respond by the same amount to changes
between the two extreme neighborhood compositions.

\subsection{Snowdrift game}
The snowdrift game is characterized by the payoff vector $(R,S,T,P)=\left(b- c/2,b-c,b,0\right)$ with $b>c>0$. Here, the average abundance of cooperation exceeds its neutral baseline if 
\[
R+S-T-P=b-\frac{3c}{2}>0,
\]
which simplifies to $b>3c/2$. Evaluating the variance correction yields
\[
(R-S)-(T-P)=\frac c2-b<0,
\]
where the inequality holds strictly because $b>c$. Thus, weak selection always decreases the stationary variance below its neutral value. In particular, when $b>3c/2$, weak selection simultaneously elevates the average abundance of cooperation and concentrates the stationary distribution more tightly around this higher mean.

Biologically, the condition $b > 3c/2$ represents the threshold where the benefit of the shared resource outweighs the risk of exploitation. In a snowdrift game, a cooperator paired with a defector must carry the entire workload alone, but still receives the benefit ($b-c$). When the benefit is sufficiently high ($b > 3c/2$), the payoffs from both mutual cooperation and one-sided cooperation are high enough to consistently satisfy more cooperators' aspirations, allowing them to revise their strategy less often compared to defectors. 

The negative variance correction has a different interpretation. A cooperator's payoff decreases by
\(R-S=c/2\) when its neighborhood changes from all cooperative to all
defecting, whereas the corresponding decrease for a defector is
\(T-P=b\). Since \(b>c>0\), the defector payoff changes more strongly between the two
extreme neighborhood compositions, producing a negative first-order variance correction.

\subsection{Stag-hunt game}
Next, we consider the stag-hunt game, parameterized by $(R,S,T,P)=(b-c,-c,0,0)$ with $b>c>0$. Cooperation is favored over defection at first order if and only if
\[
R+S-T-P=b-2c>0,
\]
or equivalently, $b>2c$. The corresponding variance correction is given by
\[
(R-S)-(T-P)=b>0,
\]
showing that weak selection strictly increases the stationary variance above its neutral value. In this scenario, cooperator payoffs depend heavily on whether their immediate neighbors cooperate, while defector payoffs remain constant at zero. This structural asymmetry drives wider strategic fluctuations under weak selection.

Biologically, the condition \(b>2c\) represents the threshold at which the benefit of successful coordination is large enough to overcome the risk of  failed coordination. In a stag-hunt game, defectors receive the safe payoff \(0\), independently of their local environment. Thus, if an individual remains a defector for some time, its aspiration level is pulled toward \(0\), its surplus becomes close to zero, and its switching probability remains close to the neutral value \(1/2\). Cooperators behave differently: mutual cooperation gives the high payoff \(b-c\), which can keep payoffs above aspiration and reduce switching, whereas failed coordination gives the payoff \(-c\), creating dissatisfaction and promoting switching. The condition \(b>2c\) means that, under the neutral local environment, the satisfaction gained from successful coordination is large enough to compensate for the dissatisfaction caused by failed coordination. Hence cooperators are, on average, less likely than defectors to abandon their strategy under weak selection, and the stationary abundance of cooperation is shifted above \(1/2\).

The strictly positive variance correction reflects an asymmetry in
payoff sensitivity to neighborhood composition. A defector's payoff
is unchanged when its neighborhood changes from all cooperative to
all defecting, whereas a cooperator's payoff decreases by \(b\).
Thus, cooperator payoffs are more sensitive to neighborhood
composition than defector payoffs, which increases the stationary
variance of the cooperator frequency above its neutral value at first
order.
\subsection{Harmony games}
Finally, in harmony games, cooperation strictly dominates defection, meaning
\(R>T\) and \(S>P\). This dominance ensures that
\[
R+S>T+P,
\]
so the average abundance of cooperation always exceeds its neutral baseline under weak selection. The interpretation is direct: for any local environment, a cooperator receives a higher payoff than a defector facing the same environment.
This means that cooperators are, on average, more likely to experience payoffs above
their aspiration and less likely to revise their strategy. This creates a
systematic drift toward higher stationary abundance of cooperation.

However, strategic dominance alone does not determine the direction of the
variance correction. The first-order variance increases, decreases, or remains
unchanged according to
\[
R-S \gtrless T-P.
\]
This comparison measures which strategy is more sensitive to local neighborhood
composition. If \(R-S>T-P\), cooperator satisfaction varies more strongly with
the local environment, and aspiration updating amplifies fluctuations in the
cooperator frequency. If \(R-S<T-P\), defector satisfaction is more sensitive to
the local environment, producing a stabilizing feedback that reduces stationary
fluctuations. If \(R-S=T-P\), both strategies respond equally to local
composition at first order, and the stationary variance remains unchanged to
first order. Thus, even when cooperation is always favored in abundance, the
spread of the stationary distribution depends on the
payoff-sensitivity contrast, not on dominance alone.

Table~\ref{tab:applications-mean-variance} summarizes these analytical findings across all four game classes.

\begin{table}[htbp]
\centering
\small
\renewcommand{\arraystretch}{1.2}
\begin{tabular}{@{}llll@{}}
\toprule
Game
& \((R,S,T,P)\)
& Abundance contrast
& Variance contrast
\\
\midrule
Donation
& \((b-c,-c,b,0)\)
& \(-2c<0\)
& \(0\)
\\
Snowdrift
& \(\left(b-c/2,b-c,b,0\right)\)
& \(b-3c/2\)
& \(c/2-b<0\)
\\
Stag hunt
& \((b-c,-c,0,0)\)
& \(b-2c\)
& \(b>0\)
\\
Harmony
& \(R>T,\;S>P\)
& \(R+S-T-P>0\)
& \((R-S)-(T-P)\)
\\
\bottomrule
\end{tabular}
\caption{
First-order stationary abundance and variance corrections for standard
pairwise games. The abundance and variance corrections have the signs
of the contrasts in the final two columns. A zero entry means only
that the first-order term vanishes.
}
\label{tab:applications-mean-variance}
\end{table}


\section{Monte Carlo simulations}\label{sec:simulations}

The simulations serve four purposes. First, on random regular
graphs, we check the first-order predictions for the stationary mean and
scaled variance across degrees $k$ and population sizes $N$, and track how
well the approximations hold as $\beta$ grows. Second, under averaged payoffs,
we compare four connected graph families with different degree distributions
to test whether their behavior near neutrality is consistent with the
topology-independent analytical results, and to observe where topology matters
at larger $\beta$. Third, we contrast averaged and accumulated payoffs to show
how the payoff-accounting convention and degree heterogeneity shape the
stationary moments; the accumulated-payoff results are discussed descriptively
since that process lies outside the scope of the theorems. Fourth, we vary the
learning rate $\rho$ over a wider selection range to examine how it affects
the stationary process beyond weak selection. All comparisons are visual and
qualitative; they do not estimate derivatives at $\beta=0$.

\subsection{Simulation setup}\label{ssec:mc-setup}
For the weak-selection regime,  we use connected random \(k\)-regular graphs. To test degree dependence, we fix \(N=50\) and take \(k\in\{4,6,8\}\). To test population-size dependence, we fix \(k=4\) and take \(N\in\{20,50,100\}\).

To examine the effect of graph structure, we consider four
connected graph families with \(N=50\) and comparable mean degrees. In the
random \(4\)-regular graph, every vertex has exactly four neighbors. The
Erd\H{o}s--R\'enyi graph is generated with edge probability
\(p=4/(N-1)\), giving expected degree four before conditioning on
connectedness. The Barab\'asi--Albert graph is generated with attachment
parameter \(m=2\), so each new vertex connects preferentially to two existing
vertices, producing a heterogeneous degree distribution with mean degree close
to four. Finally, the wheel graph consists of a cycle of \(N-1\) peripheral
vertices connected to a central hub; the hub has degree \(N-1\), each
peripheral vertex has degree \(3\), and the mean degree is
\(4(N-1)/N\). Thus, the four graph families have similar average connectivity
but substantially different degree distributions and structural
organization.

Each elementary update follows the microscopic rule in Section~2. One vertex
is selected uniformly. Its current payoff and payoff--aspiration surplus are
computed, its strategy is switched with the Fermi probability, and its
aspiration is then updated using the payoff obtained before switching. Initial
strategies are independent, with \(C\) and \(D\) equally likely. Under averaged
payoffs, every aspiration initially equals
\[
\overline{\pi}=\frac{R+S+T+P}{4}.
\]
Under accumulated payoffs, the initial aspiration of vertex \(i\) is sampled
uniformly from $[d_i\pi_-,d_i\pi_+]$, where \(d_i\) is its degree.

Unless stated otherwise, \(\rho=0.5\) and
\[
\beta\in\{0,0.025,\ldots,0.5\}.
\]
Each run consists of \(\mathcal T_{\mathrm{burn}}=2000N\) burn-in updates,
followed by \(\mathcal T_{\mathrm{meas}}=10000N\) measured updates. We use
eight independent replications for every parameter configuration. Graph
realizations are independent across replications and, whenever the graph
parameters are unchanged, are paired across parameter values. Markers show
averages over the eight replications.

\subsection{Weak-selection benchmark}
\label{ssec:mc-benchmark}

\begin{figure}[H]
\centering
\includegraphics[width=\textwidth]{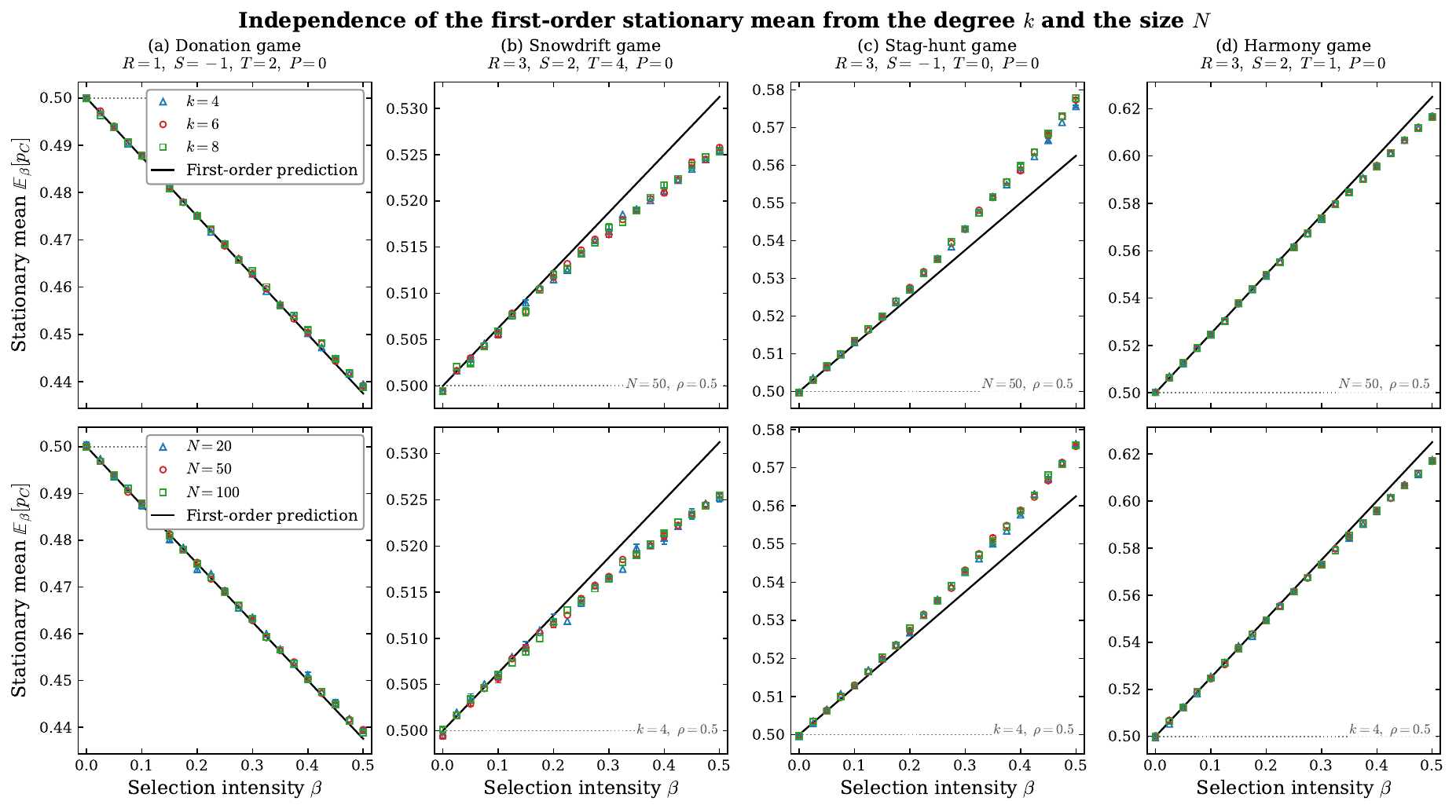}
\caption{Stationary mean under averaged payoffs on connected random regular
graphs with \(\rho=0.5\). Columns correspond to the donation, snowdrift,
stag-hunt, and harmony games. The top row fixes \(N=50\) and compares
\(k\in\{4,6,8\}\); the bottom row fixes \(k=4\) and compares
\(N\in\{20,50,100\}\). Markers show averages over eight replications. Solid black lines are the first-order
predictions, and dotted gray lines mark the neutral mean \(1/2\). Near neutrality, the simulated values remain visually
close and are consistent with the predicted independence from \(k\)
and \(N\).}
\label{fig:weak-mean}
\end{figure}

Figure~\ref{fig:weak-mean} compares the simulated
stationary abundance with its first-order approximation. The top row
fixes \(N=50\) and varies \(k\), whereas the bottom row fixes \(k=4\) and
varies \(N\). At the smallest displayed values of \(\beta\), the
curves for different values of \(k\) and \(N\) remain close to one
another and to the first-order lines. This behavior is visually
consistent with the analytical result that the first-order abundance
coefficient is independent of degree and population size.

The linear approximation becomes less accurate as
\(\beta\) increases. The donation-game mean remains close to the
prediction over the displayed range. The snowdrift and harmony means
fall below their linear approximations, whereas the stag-hunt mean
rises above its approximation.

\begin{figure}[H]
\centering
\includegraphics[width=\textwidth]{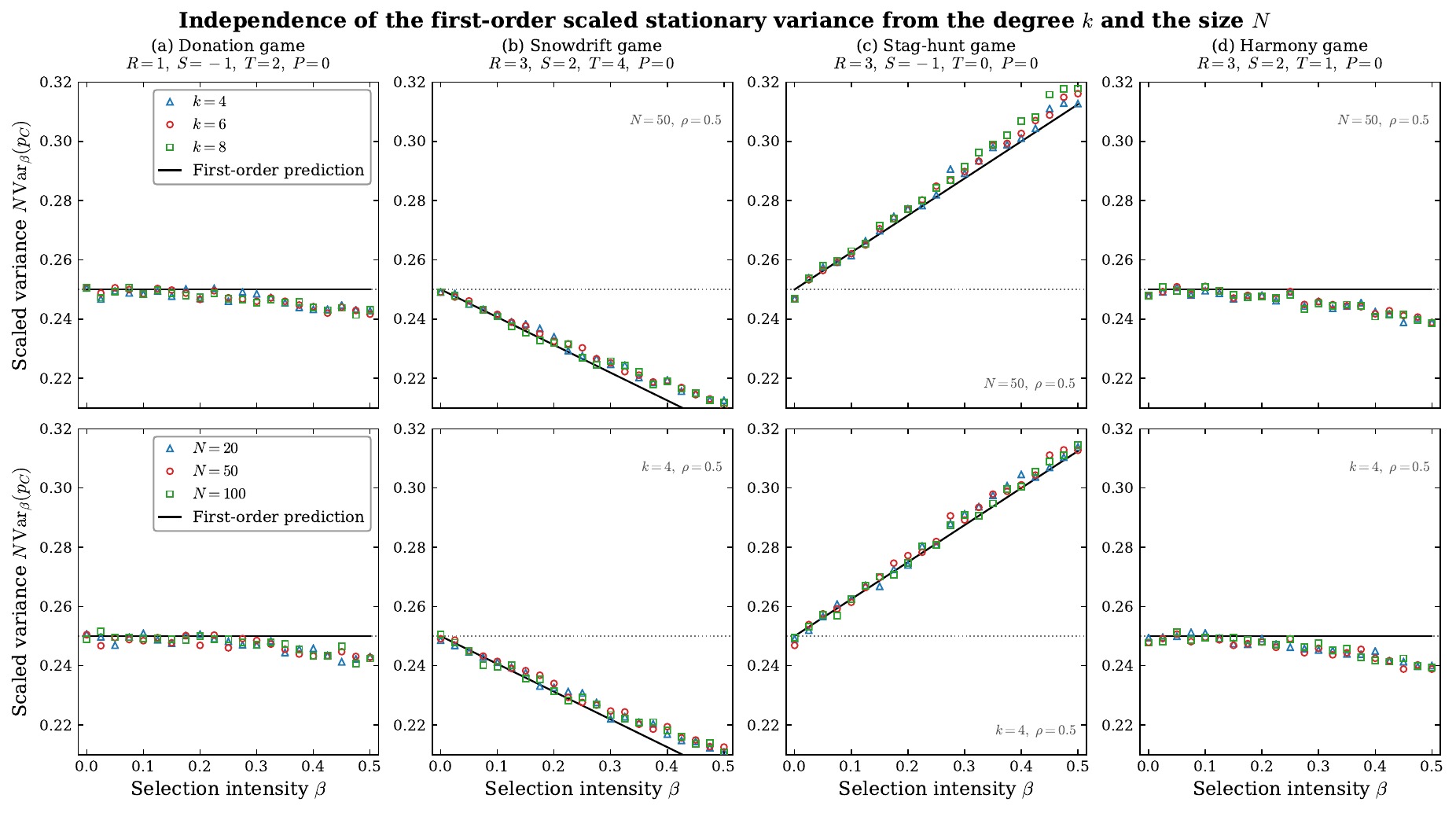}
\caption{Scaled stationary variance under averaged payoffs on connected random
regular graphs with \(\rho=0.5\). The top row fixes \(N=50\) and compares
\(k\in\{4,6,8\}\); the bottom row fixes \(k=4\) and compares
\(N\in\{20,50,100\}\). Markers show averages over eight replications. Solid
black lines are the first-order predictions, and dotted gray lines mark the
neutral value \(1/4\). The qualitative near-neutral trends are visually
consistent with the predicted coefficients and their independence
from \(k\) and \(N\).}
\label{fig:weak-variance}
\end{figure}

Figure~\ref{fig:weak-variance} gives the corresponding comparison for the scaled
stationary variance. The analytical expansion is
\[
N\operatorname{Var}_{\beta}(p_C)=\frac{1}{4}+\frac{\beta}{32}\bigl[(R-S)-(T-P)\bigr]+o(\beta).
\]
It predicts a negative first-order coefficient in the snowdrift game,
a positive coefficient in the stag-hunt game, and a zero coefficient
in the donation and harmony games. The simulated values show the
corresponding qualitative trends near neutrality. In the donation and
harmony games, they remain close to the neutral value at the smallest
values of \(\beta\) and decrease mildly as \(\beta\) increases, which
is compatible with a zero first-order coefficient and nonzero
higher-order terms. The near-neutral overlap among the curves for
different values of \(k\) and \(N\) is visually consistent with the
analytical prediction.

\subsection{Graph topology and payoff accounting}
\label{ssec:mc-topology}

\begin{figure}[H]
\centering
\includegraphics[width=\textwidth]{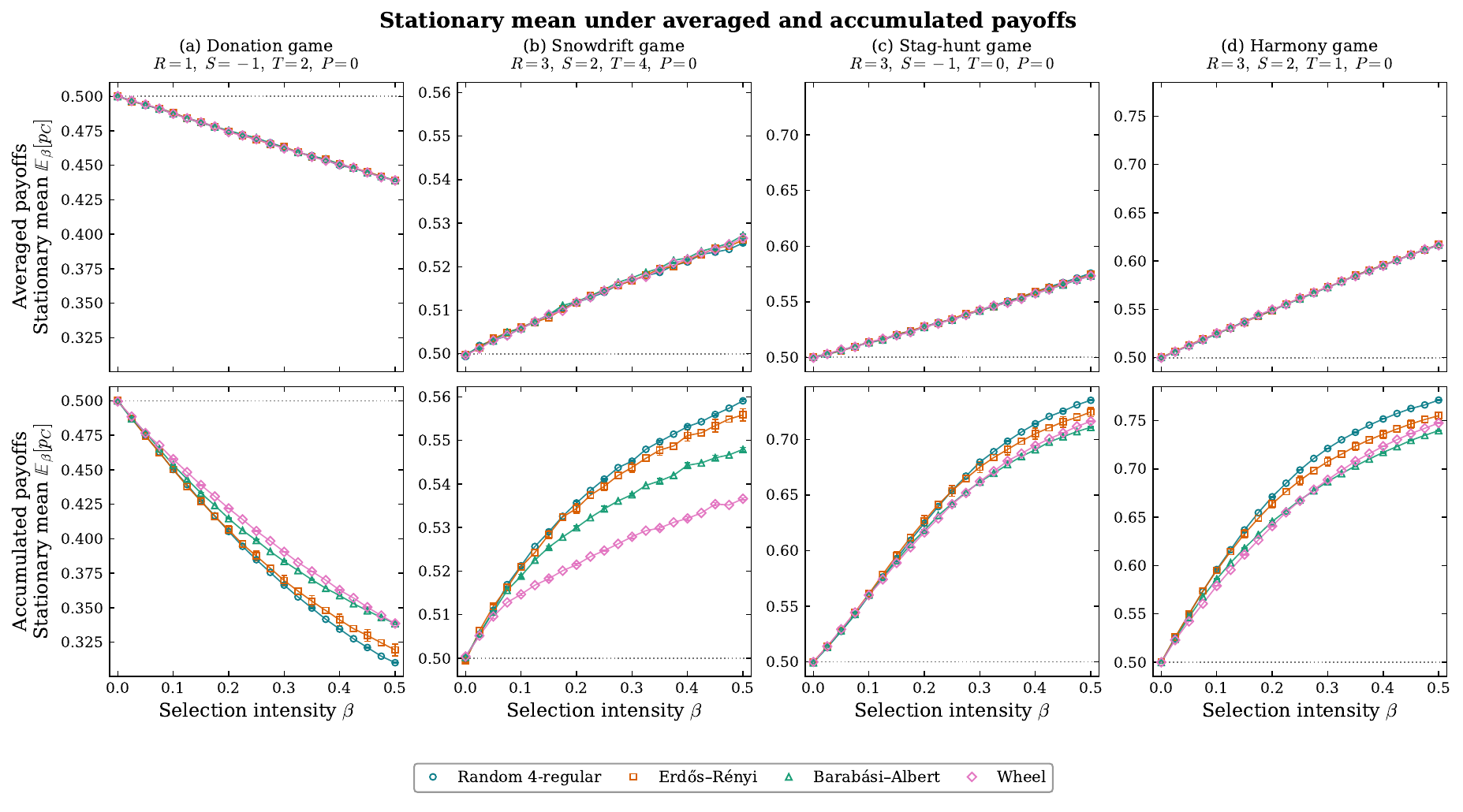}
\caption{Stationary mean on four connected graph families with \(N=50\) and
\(\rho=0.5\). Columns correspond to the four games. The top row uses averaged
payoffs, and the bottom row uses accumulated payoffs. Markers show averages
over eight replications. Dotted gray
lines mark the neutral mean \(1/2\). Under averaged payoffs, the four graph
families remain close over the displayed range. Accumulated payoffs strengthen
the response to selection and produce clear topology-dependent differences at
larger \(\beta\).}
\label{fig:topology-mean}

\end{figure}

Figure~\ref{fig:topology-mean} compares the stationary
mean on the four graph families. Under averaged payoffs, shown in the
top row, the four curves remain close throughout the displayed range.
Near neutrality, they follow the same qualitative trend: cooperation
decreases in the donation game and increases in the other three
games. This behavior is visually consistent with the
topology-independent first-order result in
Theorem~\ref{thm:exact_first_order_abundance}.

Accumulated payoffs, shown in the bottom row, produce
a substantially stronger response to selection. The curves remain
relatively close near neutrality. The similar average degrees of the
four graph families may help explain this pattern, but this
interpretation is only heuristic: no first-order formula for
accumulated payoffs on irregular graphs is derived here. As \(\beta\)
increases, the curves separate, showing numerically that the degree
distribution and topology affect the finite-selection behavior. At
\(\beta=0.5\), the random \(4\)-regular graph exhibits the largest
response in magnitude: it has the lowest cooperation level in the
donation game and the highest level in the snowdrift, stag-hunt, and
harmony games. The Barab\'asi--Albert and wheel graphs generally show
weaker responses, although their relative ordering depends on the
game. Thus, the accumulated-payoff results cannot be summarized by
the average degree alone.

\begin{figure}[H]
\centering
\includegraphics[width=\textwidth]{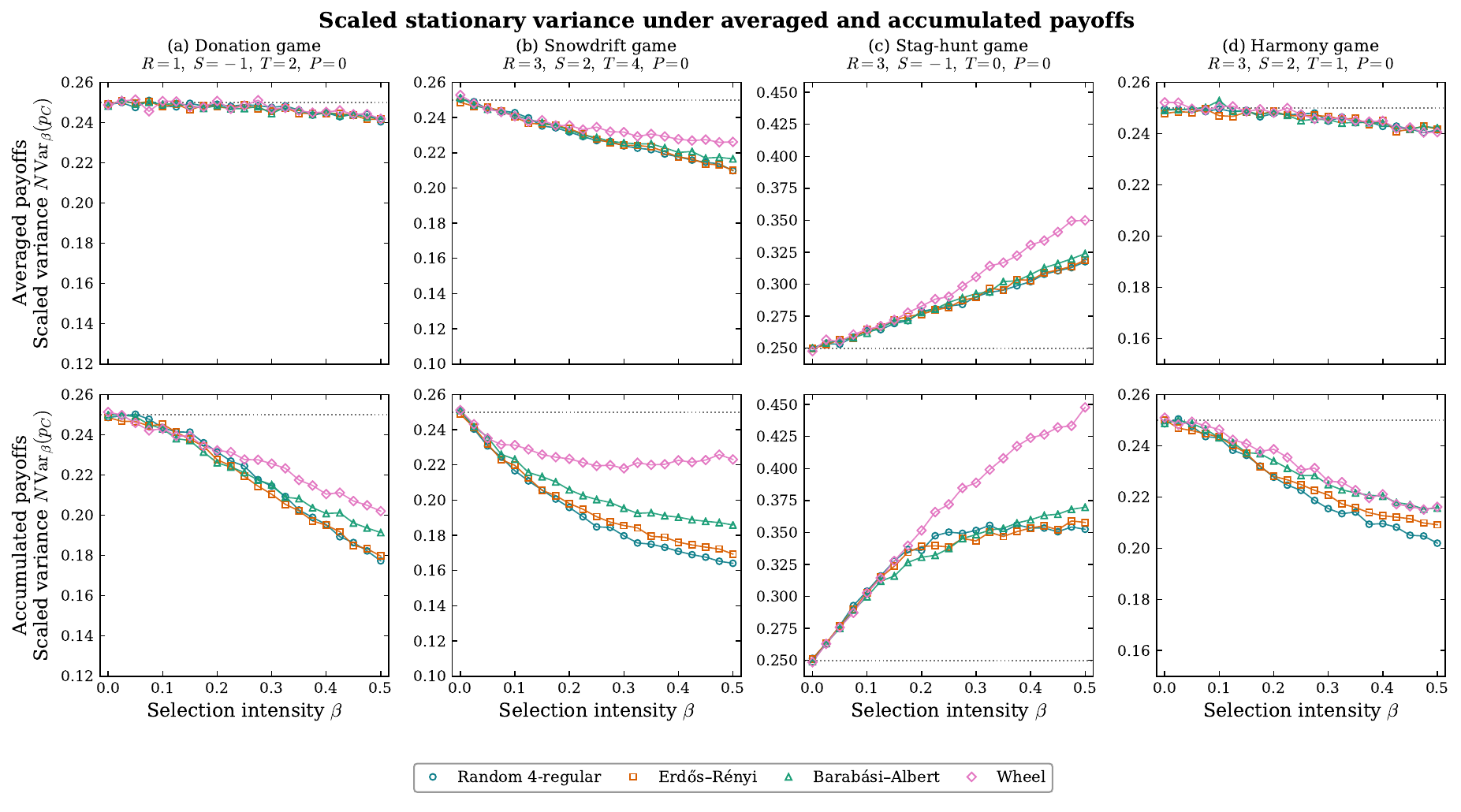}
\caption{Scaled stationary variance on four connected graph families with
\(N=50\) and \(\rho=0.5\). Columns correspond to the four games. The top row
uses averaged payoffs, and the bottom row uses accumulated payoffs. Markers
show averages over eight replications, and dotted gray lines mark the neutral
value \(1/4\). With averaged payoffs, the graph families remain visually close near neutrality
but separate at larger \(\beta\) in the snowdrift and stag-hunt games.
Accumulated payoffs produce substantially larger topology-dependent
differences.}
\label{fig:topology-variance}
\end{figure}

Figure~\ref{fig:topology-variance} shows the scaled
stationary variance. Theorem~\ref{thm:variance-first-order} predicts a
zero first-order coefficient in the donation and harmony games, a
negative coefficient in the snowdrift game, and a positive coefficient
in the stag-hunt game. Under averaged payoffs, the simulated curves
show the corresponding qualitative near-neutral trends and remain
close across the four graph families. At larger values of \(\beta\),
topology-dependent differences become visible in the snowdrift and
stag-hunt games. In both cases, the wheel graph has the largest
stationary variance among the four structures.

Under accumulated payoffs, both the response to selection and the separation
among graph families are stronger. The wheel graph has the largest, or nearly
largest, scaled variance in every game. Its variance decreases much less in the
snowdrift game and increases much more in the stag-hunt game than on the other
graphs. These results show directly that the topology-independent first-order
result for averaged payoffs does not imply topology-independent stationary
dispersion under accumulated payoffs or stronger selection. These accumulated-payoff results are interpreted
descriptively and are not compared with an analytical first-order
benchmark.

\subsection{The learning rate beyond weak selection}
\label{ssec:mc-rho-strong}

The absence of \(\rho\) from the first-order coefficients in the expansions of $\mathbb E_\beta[p_C]$ and $\operatorname{Var}_{\beta}(p_C)$ does not imply that the stationary process is independent of the aspiration learning rate. To examine this dependence, we use random \(4\)-regular graphs with averaged payoffs, \(N=50\), and 
\[
\rho\in\{0.1,0.5,1\}.
\]
The weak-selection grid up to \(\beta=0.5\) is retained, and a coarser grid is
used from \(\beta=0.75\) to \(\beta=4\). All other simulation settings remain
unchanged.

\begin{figure}[H]
\centering
\includegraphics[width=\textwidth]{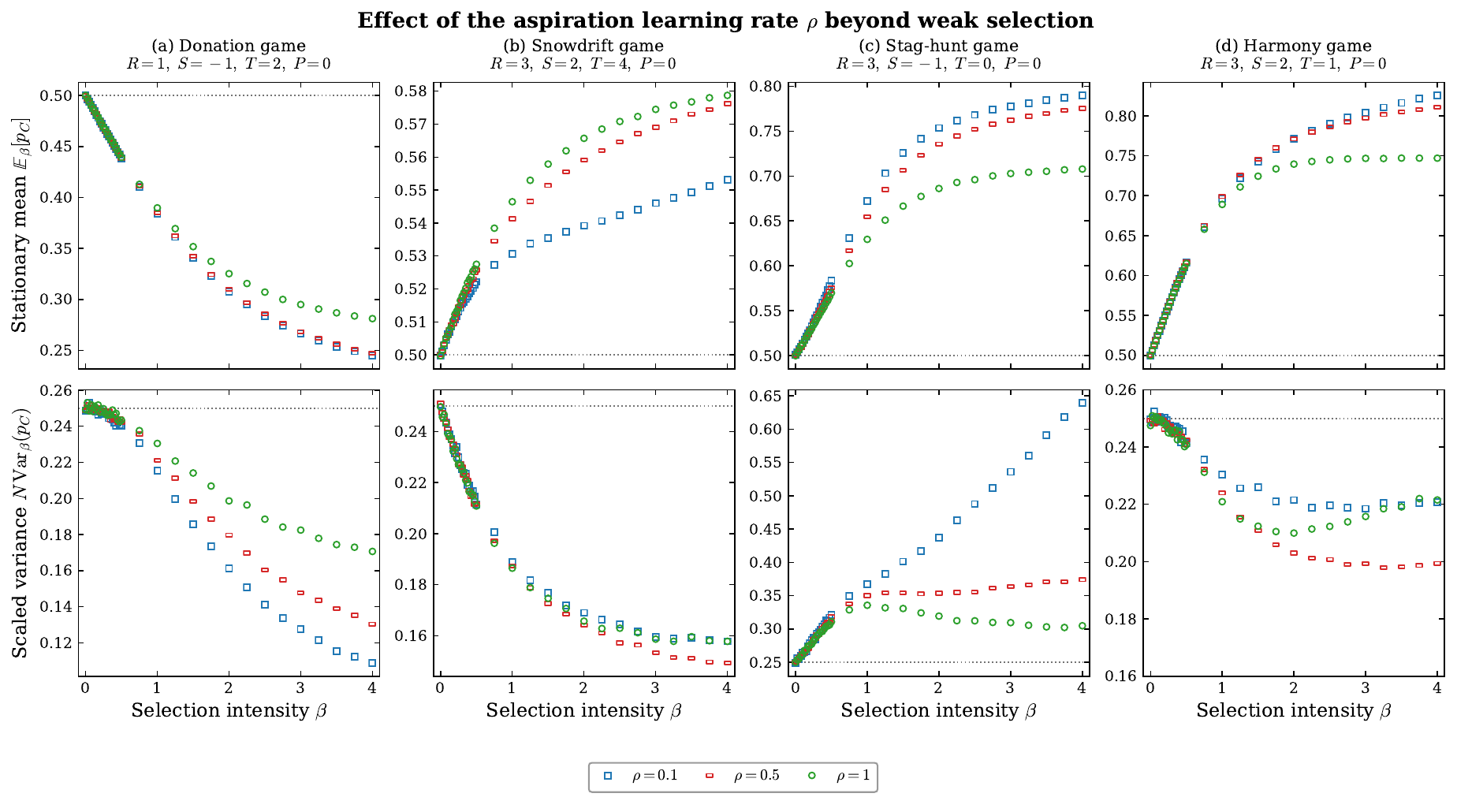}

\caption{Effect of the aspiration learning rate beyond weak selection on
random \(4\)-regular graphs with averaged payoffs and \(N=50\). The top row
shows the stationary mean, and the bottom row shows the scaled stationary
variance. Blue squares, red rectangles, and green circles correspond to
\(\rho=0.1\), \(0.5\), and \(1\), respectively. Each point is a mean over eight replications; dotted gray lines
mark the neutral values. The three curves nearly coincide for $\beta\leq0.5$
and diverge as selection intensifies, providing numerical evidence that \(\rho\)-independence
is a first-order result and does not hold beyond weak selection.}
\label{fig:rho-strong}
\end{figure}

Figure~\ref{fig:rho-strong} shows that the three mean curves are close for weak
selection but separate progressively as \(\beta\) increases. At \(\beta=4\),
the stationary mean ranges approximately from \(0.245\) to \(0.281\) in the
donation game, from \(0.553\) to \(0.579\) in the snowdrift game, from
\(0.708\) to \(0.790\) in the stag-hunt game, and from \(0.747\) to \(0.825\)
in the harmony game. Increasing \(\rho\) raises cooperation in the donation
and snowdrift games but lowers it in the stag-hunt and harmony games over the
strong-selection range considered here.

The scaled variance also depends on \(\rho\), but the direction of this effect depends on the game. In the donation game, the variance is largest for \(\rho=1\) and smallest for \(\rho=0.1\). In the stag-hunt game, this ordering is reversed, and slow aspiration learning widens the stationary distribution. In the snowdrift and harmony games, the dependence on \(\rho\) is nonmonotone, with the intermediate learning rate $\rho=0.5$ producing the smallest variance under strong selection. Hence, \(\rho\) drops out of the analytical coefficients only to first order in \(\beta\). It is not absent from the stationary process itself.

\section{\label{sec:discussion}Discussion}

In this paper, we have studied adaptive aspiration dynamics on
finite connected graphs with degree-normalized average
payoffs. Each individual has a strategy and an aspiration level, where the
latter acts as an internal benchmark for satisfactory performance. Aspirations
are updated from experienced payoffs, and strategy revision follows a Fermi
self-switching rule, where an individual compares its realized payoff with its
aspiration level. A satisfied individual is more likely to maintain its current
strategy, whereas a dissatisfied individual is more likely to switch. In both
cases, the individual will update its aspiration level by including its current
payoff in the revision. Under such a rule, the evolutionary process is
non-absorbing since even a monomorphic population can experience
self-switching. Thus, the natural long-run quantity used to quantify the
prevalence of cooperation is stationary abundance.

Because the stochastic process features continuous aspiration variables,
establishing the existence and uniqueness of the stationary distribution is
not straightforward. By applying an argument based on coupling Markov chains,
we have proven that the process restricted to $\mathcal{E}$ admits a unique stationary distribution. This
proof ensures that the finite-population stationary abundance is well defined
for any selection intensity \(0\leq\beta<\infty\), any population size
\(N\geq2\), and any finite connected graph. Consequently, our
long-run abundance results are derived from the stochastic process itself,
rather than from a deterministic closure or a large-population approximation.

Our first analytical result is the first-order expansion for the average
abundance of cooperation under the weak selection regime, where
\[
\mathbb E_\beta\left[p_C\right]=\frac12+\frac{\beta}{16}(R+S-T-P)+o(\beta).
\]
This formula shows that weak selection increases the abundance of cooperation
above its value under neutrality if \(R+S>T+P\). This outcome can be
understood through the following intuition. In the neutral local environment,
a neighbor is cooperative with probability \(1/2\). Consequently, a cooperator
earns an average payoff of \((R+S)/2\), while a defector earns an average payoff
of \((T+P)/2\). If \(R+S>T+P\), cooperation yields a higher payoff than
defection in the neutral local environment. This means that cooperators are
more satisfied on average and abandon their strategy less often than
defectors. This persistence effect shifts the stationary abundance of
cooperation above its neutral value. Conversely, if \(R+S<T+P\), defection
yields higher average satisfaction and cooperators abandon their strategy more
often than defectors. Thus, the average abundance of cooperation falls below
its value under neutrality. If \(R+S=T+P\), the first-order
shift vanishes, and higher-order terms may determine the direction of the
stationary shift.

Our second analytical result is the expansion of the variance of the abundance
of cooperation to first order, which is given by
\[
\operatorname{Var}_{\beta}(p_C)=\frac{1}{4N}+\frac{\beta}{32N}\bigl[(R-S)-(T-P)\bigr]+o(\beta).
\]
This shows that the first-order mean shift and the first-order variance shift
measure different aspects of the game. The variance shift is controlled by the
payoff-range contrast \((R-S)-(T-P)\), which compares how strongly the payoff
of each strategy responds to changes in local neighborhood composition.
The quantity \(R-S\) is the change in a cooperator's average
payoff as the fraction of cooperative neighbors increases from \(0\) to \(1\),
while \(T-P\) is the analogous change for a defector. If \(R-S>T-P\), then weak
selection increases the stationary variance of \(p_C\) above its value under
neutrality. If \(R-S<T-P\), then weak selection decreases it. Note that a
positive variance correction indicates larger stationary fluctuations in
cooperation, whereas a negative correction indicates a more concentrated
stationary distribution.

The classical games considered in the manuscript illustrate that the mean and
variance effects need not have the same sign. In the donation game, the mean
shifts against cooperation, but the variance correction vanishes because both
strategies have the same payoff range. In the snowdrift game, weak selection
can increase the stationary mean of cooperation while decreasing the
stationary variance, making the distribution more concentrated. In the
stag-hunt game, weak selection can increase both the mean and the variance,
reflecting the coordination character of the game. In the harmony game,
cooperation is favored in mean abundance, but the sign of the variance
correction still depends on whether \(R-S\) is larger or smaller than \(T-P\).
This confirms that the variance correction should not be viewed as another
version of the abundance condition. It gives complementary information: the
abundance expansion describes the direction of the stationary mean, while the
variance expansion describes the first-order change in the spread of the
stationary cooperator frequency.

The aspiration learning rate \(\rho\) and the topology of the graph do not enter the
first-order abundance condition or the first-order variance correction. This
absence does not mean that the learning process or graph structure is
irrelevant. At neutrality, strategies are independent along
edges, and the usual effects of assortment and reproductive competition are
absent because the update rule does not involve social learning or
reproduction. The strategy-specific aspiration contributions cancel in the
stationary drift balances, while average-payoff normalization removes the
explicit dependence on vertex degree.

The learning rate determines how quickly aspirations adjust to
experienced payoffs. Large values of $\rho$ produce rapid adjustment
and short memory, whereas small values produce slow adjustment and
long memory. The simulations in Figure~\ref{fig:rho-strong} provide
numerical evidence that the estimated stationary mean and scaled
variance depend on \(\rho\) when selection is no longer weak. The
direction and magnitude of this dependence vary across games.
This means that the absence of \(\rho\) from the analytical coefficients is
a first-order weak-selection result, not a general invariance property
of the model.

Similarly, if accumulated payoffs were used instead of average payoffs,
then on a \(k\)-regular graph the first-order abundance
coefficient would be multiplied by \(k\), though the sign condition would
remain the same. On an irregular
graph, the scale depends on the focal vertex, and the process cannot
be reduced to the averaged-payoff process through a single global
rescaling of \(\beta\). Ultimately, both the learning rate and graph
structure can affect features outside the scope of first-order weak selection,
including transient dynamics, relaxation and mixing times, finite-selection
behavior, and higher-order corrections in \(\beta\).

The novelty of the present work is not the algebraic inequality
\(R+S>T+P\) itself. Related weak selection criteria are already known in
aspiration dynamics, especially in models where aspiration levels are fixed
parameters or externally prescribed constants
\citep{DuWuAltrockWang2014,DuWuWang2015,WuZhou2018,ZhouWuDuWang2021}.
Such fixed aspiration formulations are mathematically useful, but they do not
capture the fact that aspirations are shaped by the payoffs actually
experienced by individuals. In biological and behavioral settings, a more
natural assumption is that satisfaction thresholds evolve together with the
population state, as in
adaptive aspiration-learning and satisficing models
\citep{KarandikarMookherjeeRayVegaRedondo1998,MacyFlache2002,
MasudaNakamura2011}. Here, aspirations are therefore treated as endogenous
state variables. They form an additional component of the evolutionary
process and depend on the history of realized payoffs through the recursive
learning rule.

The contribution is also methodological. The analysis is carried out directly
on the finite stochastic graph process, for every finite population size
\(N\) and every finite connected graph under
degree-normalized average payoffs. No large population limit, pair
approximation, mean-field closure, or
locally tree-like assumption is used. This is important
because many graph-based weak selection calculations rely on local tree
approximations or asymptotic arguments in which short cycles and finite-size
effects are suppressed. By contrast, the formulas obtained here are stationary
finite-population identities for the exact microscopic process.

Finally, the variance calculation provides information that is usually absent
from weak-selection abundance criteria. Most aspiration-dynamics studies
focus on outcome measures such as stationary abundance
\citep{DuWuAltrockWang2014,DuWuWang2015,WuZhou2018,ZhouWuDuWang2021},
fixation probabilities \citep{Kroumi2021,Kroumi2026}, or
evolutionary performance and stability
\citep{Posch1999WSLS}. The present analysis also gives the first-order
correction to the stationary variance of the cooperator frequency. This shows
that the mean shift and the fluctuation shift are controlled by different
payoff contrasts: \(R+S-T-P\) governs the stationary abundance, whereas
\((R-S)-(T-P)\) governs the first-order change in stationary dispersion.
Thus, the model distinguishes between selection favoring cooperation on
average and selection making cooperation levels more or less variable.


\section*{Acknowledgments}
The author sincerely thanks the two anonymous reviewers for their
careful reading and constructive comments. Their suggestions
substantially improved the manuscript.

\section*{Data Availability Statement}
Data will be made available on request.

\appendix
\section{Proof of Theorem~\ref{thm:stationary_distribution}}\label{appendixA}

The proof is organized as follows. We first prove existence of an
invariant probability measure in
Lemma~\ref{lem:existence_invariant_measure}, using the Feller property
of the process on the compact state space. We then recall the
asymptotic-coupling uniqueness criterion in
Lemma~\ref{lem:hms_uniqueness}. Next,
Lemma~\ref{lem:finite-reset-block} constructs a finite reset block
which brings any two copies of the chain close to the all-defector
state with positive probability. This reset block is used in
Proposition~\ref{lem:reset_asymptotic_coupling} to construct an asymptotic coupling, which gives uniqueness
of the invariant measure. Finally, the proof of
Theorem~\ref{thm:stationary_distribution} combines existence and
uniqueness with a weak-convergence argument to obtain continuity of
the invariant measure at neutrality.

\begin{lemma}\label{lem:existence_invariant_measure}
Let \(G=(V,E)\) be a finite connected graph with \(N=|V|\ge2\).
Assume \(0<\rho\le1\) and \(0\le\beta<\infty\). Then the aspiration-learning process restricted to
\[
\mathcal E=\{C,D\}^N\times[\pi_-,\pi_+]^N
\]
admits at least one invariant probability measure.
\end{lemma}

\begin{proof}
Let
\[
I=[\pi_-,\pi_+].
\]
The state space \(\mathcal E=\{C,D\}^N\times I^N\) is compact because \(\{C,D\}^N\) is finite and \(I^N\) is compact.

We next show that the transition kernel is Feller. For a fixed strategy profile \(\boldsymbol{\mathcal{S}}\), the payoff \(\pi_i(\boldsymbol{\mathcal{S}})\) is fixed, the aspiration update map
\[
A_i\mapsto (1-\rho)A_i+\rho\pi_i(\boldsymbol{\mathcal{S}})
\]
is affine, and the switching probability
\[
\sigma_i(\boldsymbol{\mathcal{S}},\mathbf A)
=
\frac{1}{1+\exp\{\beta(\pi_i(\boldsymbol{\mathcal{S}})-A_i)\}}
\]
is continuous in \(A_i\). Since the selected vertex and the switch/no-switch outcome range over finitely many possibilities, the one-step transition operator maps continuous functions on \(\mathcal E\) to continuous functions on \(\mathcal E\). Hence the process is a Feller Markov chain on a compact metric space.

Because \(\mathcal E\) is compact, the Cesàro averages of the transition probabilities are tight for every initial condition. The Feller property therefore implies the existence of an invariant probability measure by \citet[Theorem~12.1.2(ii)]{MeynTweedie1993}.
\end{proof}


\begin{lemma}\label{lem:hms_uniqueness}
Let \(P\) be a Markov transition kernel on a Polish metric space \((E,d)\). Suppose that there exists a measurable set \(A\subseteq E\) such that every invariant probability measure of \(P\) gives mass one to \(A\). Assume furthermore that, for every \(x,y\in A\), there exists a coupling \(\Gamma_{x,y}\) of the two path laws $P^{[\infty]}(x,\cdot)$ and  $P^{[\infty]}(y,\cdot)$, such that
\[\Gamma_{x,y}\left(\left\{(x_t,y_t)_{t\ge0}:\lim_{t\to\infty}d(x_t,y_t)=0\right\}\right)>0.\]
Then \(P\) has at most one invariant probability measure.
\end{lemma}

This is the special form of Corollary~2.2 of
\citet{HairerMattinglyScheutzow2011} used below. In the present
application the couplings constructed are ordinary couplings of the
two path laws, so the absolute-continuity requirement in the general
statement is automatically satisfied.


\begin{lemma}\label{lem:finite-reset-block}
Fix \(\varepsilon>0\). For every pair of initial states \(z,\widetilde z\in\mathcal E\), there exists a finite coupling of the two chains such that, with positive probability, after finitely many steps both chains have strategy profile \(\boldsymbol{\mathcal{S}}^D=(D,\ldots,D)\) and their aspiration vectors satisfy
\[
\|\mathbf A-\widetilde{\mathbf A}\|_\infty\le 2\varepsilon.
\]
\end{lemma}

\begin{proof}
Let
\[
D_\pi=\pi_+-\pi_-.
\]
Since payoffs and aspirations both lie in \([\pi_-,\pi_+]\), we have
\[
|\pi_i(\boldsymbol{\mathcal{S}})-A_i|\le D_\pi
\]
for every state and every vertex. Hence switching and not switching both have probability bounded away from zero. More precisely, with
\[
\eta=\frac{1}{1+\exp\{\beta D_\pi\}},
\]
we have
\[
\eta\le \sigma_i(\boldsymbol{\mathcal{S}},\mathbf A)\le 1-\eta,\qquad \eta\le 1-\sigma_i(\boldsymbol{\mathcal{S}},\mathbf A)\le 1-\eta.
\]

Let
\[
z_D=(\boldsymbol{\mathcal{S}}^D,P\mathbf 1),
\]
where \(\mathbf 1=(1)_{i\in V}\). In the all-defector profile every vertex has payoff \(P\). Choose \(m\ge1\) such that
\[
(1-\rho)^mD_\pi<\varepsilon.
\]
If \(\rho=1\), this condition is immediate.

We now prescribe a finite reset block. First, each vertex is selected once and is forced to end in strategy \(D\): if it is currently \(C\), we require a switch, and if it is currently \(D\), we require no switch. Next, once all vertices are defectors, each vertex is selected \(m\) additional times and we require no switch. During these all-defector no-switch updates,
\[
A_i\mapsto (1-\rho)A_i+\rho P,
\]
and after \(m\) such updates of vertex \(i\),
\[
|A_i-P|\le (1-\rho)^mD_\pi<\varepsilon.
\]

We use the same prescribed sequence of selected vertices in the two chains. Conditional on the common selected vertex, the two switch/no-switch outcomes during the reset block are sampled independently. Each required outcome in each chain has probability at least \(\eta\), while the prescribed vertex is selected with probability \(1/N\). Thus each reset step succeeds simultaneously in the two chains with probability at least
\[
\frac{\eta^2}{N}.
\]
The whole reset block has length \(N(m+1)\), and therefore succeeds in both chains with probability at least
\[
\left(\frac{\eta^2}{N}\right)^{N(m+1)}>0.
\]
On this event, both chains have strategy profile \(\boldsymbol{\mathcal{S}}^D\), and each aspiration vector lies within \(\varepsilon\) of \(P\mathbf 1\). Consequently,
\[
\|\mathbf A-\widetilde{\mathbf A}\|_\infty\le 2\varepsilon.
\]
This proves the lemma.
\end{proof}


\begin{proposition}\label{lem:reset_asymptotic_coupling}
For every pair of initial states \(z,\widetilde z\in\mathcal E\), there exists a coupling \(\Gamma_{z,\widetilde z}\) of the two path laws $P^{[\infty]}(z,\cdot)$ and $P^{[\infty]}(\widetilde z,\cdot)$, such that
\[
\Gamma_{z,\widetilde z}\left(\left\{(z_t,\widetilde z_t)_{t\ge0}:\lim_{t\to\infty}d(z_t,\widetilde z_t)=0\right\}\right)>0.
\]
\end{proposition}

\begin{proof}
Choose \(\varepsilon>0\) so small that
\[
\frac{\beta}{4}\frac{2N\varepsilon}{\rho}<1,
\]
with the convention that this condition is automatic when \(\beta=0\). By Lemma~\ref{lem:finite-reset-block}, with positive probability the two chains reach the same all-defector strategy profile and satisfy
\[
\|\mathbf A-\widetilde{\mathbf A}\|_\infty\le 2\varepsilon.
\]
It remains to show that, after this reset, the two chains can be coupled so that they never separate again and their aspiration vectors converge to one another.

From the reset time onward, use the same selected vertex in both chains. As long as the strategy profiles agree, the payoff vectors agree as well. Conditional on the common selected vertex \(i\), couple the two switch/no-switch decisions maximally. If
\[
p=\sigma_i(\boldsymbol{\mathcal{S}},\mathbf A),\qquad q=\sigma_i(\boldsymbol{\mathcal{S}},\widetilde{\mathbf A}),
\]
then the probability that the two Bernoulli decisions disagree is exactly
\[
|p-q|.
\]
For fixed \(\boldsymbol{\mathcal{S}}\), the map
\[
a\mapsto \frac{1}{1+\exp\{\beta(\pi_i(\boldsymbol{\mathcal{S}})-a)\}}
\]
has derivative bounded by \(\beta/4\). Hence
\[
|\sigma_i(\boldsymbol{\mathcal{S}},\mathbf A)-\sigma_i(\boldsymbol{\mathcal{S}},\widetilde{\mathbf A})|\le \frac{\beta}{4}|A_i-\widetilde A_i|.
\]
Therefore, as long as the strategies still agree, the conditional probability of a disagreement at a step selecting vertex \(i\) is at most
\[
\frac{\beta}{4}|A_i-\widetilde A_i|.
\]

On the other hand, whenever the two chains make the same switch/no-switch decision, the strategy profiles remain equal. The aspiration difference at the selected vertex is multiplied by \(1-\rho\), while the other aspiration differences are unchanged:
\[
|A_i'-\widetilde A_i'|=(1-\rho)|A_i-\widetilde A_i|.
\]
Thus, conditional on the common sequence of selected vertices and on no previous disagreement,
\[
\sum_{t\ge0}|A_{I_t}(t)-\widetilde A_{I_t}(t)|\le \sum_{i=1}^N\sum_{r=0}^{\infty}(1-\rho)^r|A_i-\widetilde A_i|\le \frac{2N\varepsilon}{\rho}.
\]
By the union bound, the conditional probability of at least one future disagreement is at most
\[
\frac{\beta}{4}\sum_{t\ge0}|A_{I_t}(t)-\widetilde A_{I_t}(t)|\le \frac{\beta}{4}\frac{2N\varepsilon}{\rho}<1.
\]
Hence, with positive probability, no disagreement occurs after the reset block.

On this no-disagreement event, the strategy profiles remain identical forever. Moreover, every vertex is selected infinitely often almost surely, so each aspiration difference is repeatedly multiplied by \(1-\rho\). Therefore,
\[
\|\mathbf A(t)-\widetilde{\mathbf A}(t)\|_\infty\longrightarrow0.
\]
For the metric
\[
d((\boldsymbol{\mathcal{S}},\mathbf A),(\widetilde{\boldsymbol{\mathcal{S}}},\widetilde{\mathbf A}))=\mathbf 1_{\{\boldsymbol{\mathcal{S}}\ne \widetilde{\boldsymbol{\mathcal{S}}}\}}+\min\{1,\|\mathbf A-\widetilde{\mathbf A}\|_\infty\},
\]
this gives
\[
\lim_{t\to\infty}d\bigl((\boldsymbol{\mathcal{S}}(t),\mathbf A(t)),(\widetilde{\boldsymbol{\mathcal{S}}}(t),\widetilde{\mathbf A}(t))\bigr)=0.
\]
Since the reset block has positive probability and the subsequent no-disagreement event also has positive conditional probability, the constructed coupling assigns positive probability to the diagonal-at-infinity event. This proves the proposition.
\end{proof}


\begin{proof}[Proof of Theorem \ref{thm:stationary_distribution}]
Existence follows from Lemma~\ref{lem:existence_invariant_measure}.

Now, we will prove uniqueness. The state space \(\mathcal E\) is
compact and metrizable, hence Polish. We apply
Lemma~\ref{lem:hms_uniqueness} with \(E=A=\mathcal E\). Every invariant
probability measure gives mass one to \(\mathcal E\). Moreover,
Proposition~\ref{lem:reset_asymptotic_coupling} verifies the required asymptotic-coupling condition for every
pair of initial states \(z,\widetilde z\in\mathcal E\). Therefore the
Markov chain has at most one invariant probability measure. Since
existence has already been proved, the invariant probability measure
is unique. The stationary abundance is then well defined because
\(p_C\) is a bounded measurable function on \(\mathcal E\).

Next, we prove the continuity of the invariant measure at neutrality. Let \(\beta_n\to0\). Since \(\mathcal E\) is compact, the space \(\mathcal P(\mathcal E)\) is compact for weak convergence \citep[Theorem~5.1, p.~60]{Billingsley1999}. Hence there exist a subsequence \((\mu_N^{\beta_{n_j}})\) and a probability measure \(\nu\) on \(\mathcal E\) such that  $(\mu_N^{\beta_{n_j}})$ converges weakly to $\nu$.

Let \(P_\beta\) be the one-step transition kernel and let \(f:\mathcal E\to\mathbb R\) be a continuous bounded function. The only dependence of \(P_\beta\) on  \(\beta\) comes from the switching probabilities
\[
\sigma^\beta_i(z)=\frac{1}{1+\exp\{\beta(\pi_i(z)-A_i)\}}.
\]
Since \(\mathcal E\) is compact, the surpluses \(\pi_i(z)-A_i\) are uniformly bounded. Thus \(\sigma_i^\beta(z)\to1/2=\sigma_i^0(z)\) uniformly in \(z\) as $\beta\rightarrow0^+$. Since the number of vertices is finite and the update maps are continuous, it follows that the convergence \(P_\beta f\to P_0f\) is uniform. Using the invariance of \(\mu_N^{\beta_{n_j}}\),
\[
\int_{\mathcal E} f\,d\mu_N^{\beta_{n_j}}=\int_{\mathcal E}P_{\beta_{n_j}}f\,d\mu_N^{\beta_{n_j}}.
\]
Passing to the limit gives
\[
\int_{\mathcal E} f\,d\nu=\int_{\mathcal E}P_0f\,d\nu.
\]
Hence \(\nu\) is invariant for \(P_0\). By uniqueness of the invariant probability measure at \(\beta=0\), we have
\[
\nu=\mu_N^0.
\]
Thus every subsequential limit is \(\mu_N^0\), and therefore $(\mu_N^\beta)$ converges weakly to $\mu_N^0$ as $\beta\to0^+$. Equivalently, for every bounded continuous function \(f:\mathcal E\to\mathbb R\),
\[
\mathbb E_\beta[f]=\mathbb E_{0}[f]+o(1).
\]
This completes the proof of Theorem~\ref{thm:stationary_distribution}.

\end{proof}

\end{document}